\documentclass[structabstract]{aa}
\usepackage{graphicx}
\usepackage{txfonts}
\usepackage{natbib,twoopt}
\usepackage{epsfig}
\usepackage{graphicx}
\usepackage{tabularx}
\usepackage[applemac]{inputenc}
\usepackage{enumerate}
\usepackage{pdflscape}
\usepackage{color}
\usepackage{subfig}
\begin{document}
\title{Fossil Group Origins}

\subtitle{VII. Galaxy substructures in fossil systems}

\author{S. Zarattini \inst{1, 2, 3}, M. Girardi \inst{4,5}, J.A.L. Aguerri\inst{1,2}, W. Boschin\inst{1,2,6}, R. Barrena \inst{1, 2}, C. del Burgo\inst{7}, N. Castro-Rodriguez\inst{1, 2}, E. M. Corsini\inst{3}, E. D'Onghia\inst{8,9}, A. Kundert\inst{8}, J. M\'endez-Abreu\inst{10}, \and R. S\'anchez-Janssen\inst{11}}
%and

\institute{Instituto de Astrof\'isica de Canarias, calle Via L\'actea S/N, E-38205, La Laguna, Tenerife, Spain
\and Universidad de La Laguna, Dept. Astrof\'isica, E-38206, La Laguna, Tenerife, Spain
\and Universit\`a degli Studi di Padova, Dipartimento di Fisica e Astronomia, vicolo dell'osservatorio 3, I-35122, Padova, Italy
\and Universit\`a degli Studi di Trieste, Dipartimento di Fisica-Sezione Astronomia, via Tiepolo 11, I-34143, Trieste, Italy
\and INAF - Osservatorio Astronomico di Trieste, via Tiepolo 11, I-34143, Trieste, Italy
\and Fundaci\'on Galileo Galilei - INAF, Rambla Jos\'e Ana Fern\'andez P\'erez 7, E-38712, Bre\~na Baja, La Palma, Spain
\and Instituto Nacional de Astrof\'isica, \'Optica y Electr\'onica, Luis Enrique Erro 1, Sta. Ma. Tonantzintla, Puebla, Mexico
%\and Astronomy Department, University of Wisconsin, 475 Charter St., Madison, WI 53706, USA
\and Department of Astronomy, University of Wisconsin-Madison, 475 N. Charter St., Madison, WI 53706, USA
\and Alfred P. Sloan Fellow
\and School of Physics and Astronomy, University of St Andrews, North Haugh, St Andrews, KY16 9SS, UK
\and NRC Herzberg Institute of Astrophysics, 5071 West Saanich Road, Victoria, BC, V9E 2E7, Canada
\\ \email{stefano@iac.es}}

\date{Received}

%\abstract{}{}{}{}{}
% 5 {} token are mandatory
\abstract
{Fossil groups are expected to be the final product of galaxy merging within galaxy groups. In simulations, they are predicted to assemble their mass at high redshift. This early formation allows for the innermost $M^\ast$ galaxies to merge into a massive central galaxy. Then, they are expected to maintain their fossil status because of the few interactions with the large-scale structure. In this context, the magnitude gap between the two brightest galaxies of the system is considered a good indicator of its dynamical status. As a consequence, the systems with the largest gaps should be dynamically relaxed.}
{In order to examine the dynamical status of these systems, we systematically analyze, for the first time, the presence of galaxy substructures in a sample of 12 spectroscopically-confirmed fossil systems with redshift $z \le 0.25$.}
{We apply a number of tests in order to investigate the substructure in fossil systems in the two-dimensional space of projected positions out to $R_{200}$. Moreover, for a subsample of 5 systems with at least 30 spectroscopically-confirmed members we also analyze the substructure in the velocity and in the three-dimensional velocity-position spaces. Additionally, we look for signs of recent mergers in the regions around the central galaxies.}
{We find that an important fraction of fossil systems show substructure. The fraction depends critically on the adopted test, since each test is more sensible to a particular type of substructure.}
{Our interpretation of the results is that fossil systems are not, in general, as relaxed as expected from simulations. Our sample of 12 spectroscopically-confirmed fossil systems need to be extended in order to compute an accurate fraction, but our conclusion is that it is similar to the fraction of substructure detected in non-fossil clusters. This result points out that the magnitude gap alone is not a good indicator of the dynamical status of a system. However, the subsample of 5 FGs for which we were able to use velocities as a probe for substructures is dominated by high-mass FGs. These massive systems could have a different evolution compared to low-mass FGs, since they are expected to form via the merging of a fossil group with another group of galaxies. This merger would lengthen the relaxation time and it could be the responsible of the substructure detected in present-time massive FGs. If it is the case, only low-mass FGs are expected to be dynamically old and relaxed.}
% \keywords{}
	\authorrunning{Zarattini et al.}
	\titlerunning{Fossil Groups Origins VII}
\maketitle

%%%%%%%%%%%%%%%%%%%%%%%%%%%%%%%%%%
%%%%%%%%%%%%% SECTION 1%%% %%%%%%%%%%%%
%%%%%%%%%%%%%%%%%%%%%%%%%%%%%%%%%%
\section{Introduction}
\label{intro}

The first identification of a fossil group (FG), a type of object initially proposed by \citet{Ponman1994}, was RX-J1340.6+4018 - an apparently isolated elliptical galaxy surrounded by a X-ray halo extending to a size typical of a group-mass object. The first observational definition of FGs was given by \citet{Jones2003}. Following these authors, a group of galaxies is fossil if it has a difference of at least 2 magnitudes in the $r$-band between the two brightest member galaxies within half the virial radius. Moreover, the system must be surrounded by an extended X-ray halo of at least 10$^{42}$ $h_{50}^{-2}$ erg s$^{-1}$ to avoid the detection of isolated galaxies. There is no upper limit in the X-ray definition, thus the presence of fossil clusters is not excluded. In fact, many fossil clusters have been observed \citep{Cypriano2006,MendesdeOliveira2006,Aguerri2011,Zarattini2014}. Thus, we will refer in general to fossil systems throughout this work as pertaining to both group- and cluster-mass fossils. Nevertheless, as in the other papers of the fossil group origins (FOGO) series, we prefer to maintain the original abbreviation FG.

In the framework of the cold dark matter scenario, FGs are thought to form hierarchically at an early epoch. \citet{DOnghia2004} showed that FGs have assembled half of their mass at higher redshifts ($z>1$) than regular groups and clusters, a result confirmed also by \citet{DOnghia2005} and \citet{Dariush2010}. From $z=1$ to the current epoch, they would grow via minor mergers only, accreting only one third of the galaxies accreted by a non-fossil systems in the same amount of time \citep{vonBenda-Beckmann2008}. Nevertheless, the fraction of mass assembled by FGs is larger than that of non-FGs at any redshift \citep{Dariush2007}. All these simulations agree with the ``classic'' formation scenario, in which FGs formed earlier than non-FGs, thus having enough time to merge all the $M^\ast$ galaxy population in the single, massive central galaxy. Thus, this process would be responsible for the lack of $M^\ast$ galaxies that is observed in the luminosity function of FGs \citep{Khosroshahi2006,MendesdeOliveira2006,Aguerri2011,Khosroshahi2014,Zarattini2015}.

The merging scenario seems also to be the most valid explanation for the observed properties of the brightest group galaxies (BGGs) and intra-cluster medium (ICM). In this scenario, the main difference between fossil and non-fossil systems would be the formation time, while the evolution of both kinds of structures are similarly driven by dynamical friction and mergers. Thus, few differences are expected in the central galaxies of fossils and non-fossils, apart from the fact that, on average, the former should be more massive, as they actually are \citep{Mendez-Abreu2012}, since they have more time to accrete material. The majority of the publications on BGGs confirm that no differences are found \citep{LaBarbera2009,Cui2011,Tavasoli2011,Mendez-Abreu2012,Eigenthaler2013}, although some works claim differences in the shape of their isophotes \citep{Khosroshahi2006}.

Finally, the ICM of FGs also seems to be described by the same scaling relations of regular clusters, as it is expected if mergers are the main driver of the evolution. Some early works \citep{Proctor2011} claimed that fossils were fainter in the optical with respect to non-fossil systems, but probably these differences were due to observational biases, as shown in \citet{Voevodkin2010}, \citet{Harrison2012}, and \citet{Girardi2014}. Recently, \citet{Khosroshahi2014} suggested that, for the same X-ray luminosity, optically-selected FGs are more luminous in the optical than X-ray selected FGs. Finally, \citet{Kundert2015} have recently confirmed that the global X-ray scaling relations of fossil and non-fossil systems are consistent with one another.

If the most accepted formation scenario is correct, and FGs formed earlier than non-FGs and evolve with fewer interactions with the large-scale structure, they should be dynamically older. In this case, their galaxy population should be more relaxed, with less substructure present than in regular groups and clusters.

In fact, many studies demonstrate that clusters of galaxies are not simple relaxed structures. They are thought to evolve in a hierarchical fashion, from poor groups to rich clusters. The presence of substructures, indicative of the early phase of cluster evolution, has been observed in 30\%-70\% of clusters, as shown by optical studies \citep{Geller1982,Girardi1997,Ramella2007,Aguerri2010,Wen2013}, X-ray analysis \citep{Schuecker2001}, and gravitational lensing techniques \citep{Dahle2002, Smith2005}.

Although this aspect of clusters is generally well studied, there are very few publications in the literature on galaxy substructures in FGs. Moreover, the majority of them studied the large-scale structure around fossil systems. For example, \citet{Adami2007} analyzed the fossil system RX 1119.7+2126, finding that the system is an isolated structure. This FG, together with 1RXS J235814.4+150524 (FGS34 in the notation adopted by the FOGO series) was also studied in \citet{Adami2012}. The authors confirmed that the former is isolated, whereas the latter is embedded in a denser environment, with 10 compact structures found in an area of $\sim 50$ Mpc. \citet{Pierini2011} analyzed two FGs, also finding that one is isolated whereas the other is embedded in a more complex environment. Moreover, they claimed that FGs seem to be ``more mature'' than non-FGs of the same mass, suggesting that FGs have converted cold gas into stars in a more efficient way, and that no infalling galaxy should have been accreted in the last 3-6 Gyrs. Finally, \citet{Diaz-Gimenez2011} found, using simulations, that fossils at $z\sim 0.4$ are embedded in a denser environment than non-fossils. If the evolution of these systems is followed to $z\sim 0$, their environments are slightly under-dense.

There are very few studies of galaxy substructures within 1 or 2 $R_{200}$ in fossil systems. In the first paper of the FOGO series \citep{Aguerri2011} we analyzed in detail the fossil cluster RX J105453.3+552102 (FGS10 in the notation adopted here, see Sect. \ref{sample} for details), finding no significant sign of substructures. In fact, all the tests seem to point out that the system, located at $z=0.5$ is dynamically relaxed. \citet{Harrison2012} showed the two-dimensional density contours for 17 FGs. Nevertheless, they did not comment about the presence of substructure in their paper, with the only exception of one FG, that is noted to be unusually relaxed.

Thus, a systematic study of substructure in FGs remains to be done. For this reason, we used our sample of 12 spectroscopically-confirmed FGs with $z \le 0.25$ to investigate the presence of substructure. We apply different techniques at different radii, depending on the availability of data. For the whole sample we analyze the two-dimensional substructures out to R$_{200}$. Then, we investigate a more restricted region for a subsample of 5 FGs with at least 30 spectroscopically-confirmed members, using their velocities as a probe for substructure.

In this paper, the adopted cosmology is $H_0=70$ km s$^{-1}$ Mpc$^{-1}$, $\Omega_{\Lambda}=0.7$, and $\Omega_M =0.3$.

%%%%%%%%%%%%%%%%%%%%%%%%%%%%%%%%%%
%%%%%%%%%%%%% SECTION 2%%% %%%%%%%%%%%%
%%%%%%%%%%%%%%%%%%%%%%%%%%%%%%%%%%
\section{Sample}
\label{sample}

%\begin{landscape}
\begin{table*}
\centering
%\begin{center}
\caption{Substructure in the FGs of our sample}
\label{summary}
\begin{tabularx}{0.97\textwidth}{lcccccccccccc}
\tiny
\\
\hline
\hline
name & mass & AI & STI & weighted gap & 1D-DEDICA & $V_{\rm BGG}$ & DS & $V_{\rm grad}$ & 2D-DEDICA & VTP & $\epsilon$ & TOT+/TOT \\
\multicolumn{1}{c}{} & [M$_\odot$] & & & & & & & & & \\
\multicolumn{1}{c}{(1)}& (2) & (3) & (4) & (5) &(6) &(7) &(8) &(9) &(10) &(11) & (12) & (13) \\
\hline	
FGS02 & 1.87E+15	& N & Y & N & N & N & Y & N & Y & Y & N & 4/10\\
FGS03 & 4.20E+13	& - & - & - & - & - & - & - & N & N & N & 0/3\\
FGS14 & 5.55E+14	& N & Y & N & N & Y & N & N &Y & N & N & 3/10\\
FGS17 & -	 	& - & - & - & - & - & - & - & N & N & N & 0/3\\
FGS20 & 1.63E+14	& - & - & - & - & - & - & - & N & N & N & 0/3 \\
FGS23 & 2.86E+14	& N & N & N& Y & N & N & N & N & N & N & 1/10\\
FGS26 & 2.67E+14	& - & - & - & - & - & - & - & N & Y & N & 1/3\\
FGS27 & 6.69E+14 	& N & N & N & N & N & Y & N & Y & Y & Y & 4/10\\
%FGS28 & - 	 & - & - & - & - & - & - & - & - & - & - & Y & 1/1\\
FGS29 & 9.66E+13	& - & - & - & - & - & - & - & N & N & N & 0/3\\
FGS30 &5.57E+14	& N & Y & N & N & N & N & N & N & Y & Y & 3/10\\
FGS32 & -	 & - & - & - & - & - & - & - & N & N & N & 0/3\\
FGS34 &8.63E+13	& - & - & - & - & - & - & - & Y & N & N & 1/3\\
\hline

\end{tabularx}
%\end{center}

\tiny Note: Col. (1): System number as in \citet{Santos2007}; Col (2): System mass as in \citet{Zarattini2014}; Col (3): Asymmetry index (c.l. $\ge99\%$);

Col (4): Scale tail index (c.l. $\ge99\%$); Col (5): weighted gap (c.l. $\ge 99\%$) ; Col (6): 1D-DEDICA (c.l. $>99\%$); Col (7): Peculiar velocity of the BGG (c.l. $\ge90\%$); Col (8): DS test (c.l. $\ge 95\%$); Col (9): Velocity gradient (c.l. $\ge99\%$); Col (10): 2D-DEDICA (c.l. $\ge99\%$ and $\rho_s \ge 0.5$); Col (11): Voronoi Tessellation and Percolation (c.l. $\ge99\%$); Col (12): Ellipticity (c.l. $\ge99.7\%$); Col(13): Fraction of positive tests for each system

Y = presence of substructure; N = no substructure; - = not applicable.\\
\end{table*}
%\end{landscape}

In this work, we applied a battery of statistical tests to reveal the presence of substructure in FGs. \citet{Aguerri2010} showed that the result of the substructure analysis strongly depends on the analyzed galaxy population. The photometric completeness limit of the SDSS DR7 is $m_r=22.2$ \citep[$r$-band model magnitude,][]{Abazajian2004}, which means that we are able to reach $M_r = -19$ mag in systems with $z \le 0.25$. For this reason, here we only study the substructure for galaxies with $M_r \le -19$. The number of spectroscopically-confirmed FGs with $z \le 0.25$ in the \citet{Zarattini2014} sample is 13.
However, FGS28 is a very peculiar case, since it has only one spectroscopically-confirmed member within $R_{200}$. Also its photometric catalog is very poor and we were not able to analyze it with any of the tests presented in the next sections. For these reasons, our sample consists of 12 spectroscopically-confirmed FGs.

We applied different methods to study the presence of substructures in our sample. Some methods work in the velocity space (1D test), others in the projected space (2D test), whereas the Dressler-Schectman method \citep[][hereafter DS]{Dressler1988} works combining both spaces (3D test). As showed by \citet{Pinkney1996}, it is important to use several tests to analyze the substructure in clusters because no single test is the most sensitive for all cases. We note that the 2D tests were applied to the above-mentioned 12 FGs, whereas the 1D and 3D tests were applied to a subsample of 5 FGs with at least 30 spectroscopically-confirmed members.

%%%%%%%%%%%%%%%%%%%%%%%%%%%%%%%%%%
%%%%%%%%%%%%% SECTION 3%%% %%%%%%%%%%%%
%%%%%%%%%%%%%%%%%%%%%%%%%%%%%%%%%%
\section{Results}
\label{results}

In this section, we present the results of our statistical tests. A summary is provided in Table \ref{summary}.

\subsection{Substructure in the velocity field}
\label{1D}

Here, we present the 1D tests applied to the subsample of 5 fossil systems with $z\le 0.25$ and $N_{\rm memb} > 30$. This minimum number of cluster members was suggested by \citet{Pinkney1996}. They showed, using N-body simulations of cluster mergers, that with such a number of cluster members all the analyzed tests are able to detect substructures correctly. The systems that satisfy the previous criteria are FGS02, FGS14, FGS23, FGS27, and FGS30.

\subsubsection{Deviation from gaussianity}
\label{gauss}

The velocity distribution of galaxies in a relaxed cluster is assumed to be Gaussian. Deviations from Gaussianity are considered as an indicator of substructures.
Thus, the velocity distribution of the systems in our sample with a large number of spectroscopically confirmed members ($N_{\rm memb} > 30$) was analyzed using different estimators of Gaussianity.

The first tests we applied are the shape estimators proposed by \citet{Bird1993}. We decided to use only the asymmetry index (AI) and the scale tail index (STI), since the skewness and kurtosis are less robust and more sensitive to false-positive detections \citep{Bird1993}. Both the AI and STI measure the shape of the velocity distribution using the ordered statistics of the dataset. In particular, the AI test \citep[][]{Finch1977} evaluates the asymmetries in the velocity distribution by comparing the gaps present in the data on the left and right side of the median value \citep[$L_i$ and $R_i$ in the notation by][]{Finch1977}. In a symmetric distribution, $L_i$ and $R_i$ have the same values, whereas in case of asymmetries one of the two indicators is larger depending on the side of the asymmetry. The tail index (TI) test evaluates the asymmetries by computing the spread of the dataset at 90\% level with that of 75\% level. Then, the computed TI value is normalized to the Gaussian value, thus defining the scale tail index. More details on both procedures are given in \citet{Bird1993} and in the references therein.

We found that the AI test indicates a possible asymmetry for FGS14 (AI=1.018) and FGS27 (AI=0.991), both at a 97\% confidence level (c.l.). Nevertheless, we preferred to use a conservative c.l. $\ge 99$\% as the threshold for the detection of substructures. Thus, we considered them as not significative. The STI test suggests strong evidence of asymmetries in the distribution of FGS02 (STI=1.491), FGS14 (STI=1.741), and FGS30 (STI=0.731). These asymmetries are evaluated at c.l. $\ge$ 99\%. Thus, we can conclude that these three systems have substructures according to the STI test.
The computation of the c.l. was done by comparing our estimates of the AI and STI indexes with table 2 of \citet{Bird1993}, and also taking into account the number of member galaxies.

We also investigated the presence of gaps in the velocity distribution using the weighted gap estimator in the ordered velocity space. This method computes the differences between two contiguous velocities and weights the result for its position with respect to the center of the distribution. \citet{Beers1991} suggest to look for normalized gaps larger than 2.25. In fact, in a purely Gaussian distribution such a value appears in about 3\% of cases, independently of the sample size \citep{Wainer1978}. Moreover, we also computed the probability of finding a gap larger than 2.25 somewhere in the distribution. This cumulative probability is a more robust indicator of the importance of the measured gap, since the simple presence of a gap larger than 2.25 is not enough to claim the presence of substructures. For the fossil system FGS02 we found four gaps larger than 2.25 (2.47, 2.65, 2.77, and 3.03), and the cumulative probability of finding such gaps in a random Gaussian distribution is 2.5\%. This means that the presence of substructures is significant at 97.5\% c.l.. For FGS30 we found three gaps larger than 2.25 (2.51, 2.67, and 2.91) and the probability of finding these gaps in a random Gaussian distribution is 2\%, which leads to the presence of substructures at 98\% c.l. Nevertheless, we prefer to use a more conservative c.l. of 99\% as indicative for the presence of a real substructure, so we concluded that none of these two systems show substructure. For FGS14, FGS23 and FGS27 the cumulative probability is always smaller than 40\%, suggesting no substructure is present in these systems according to this weighted gap test.

\subsubsection{1D-DEDICA}

\begin{figure}
\centering%\begin{table}
\includegraphics[width=0.45\textwidth]{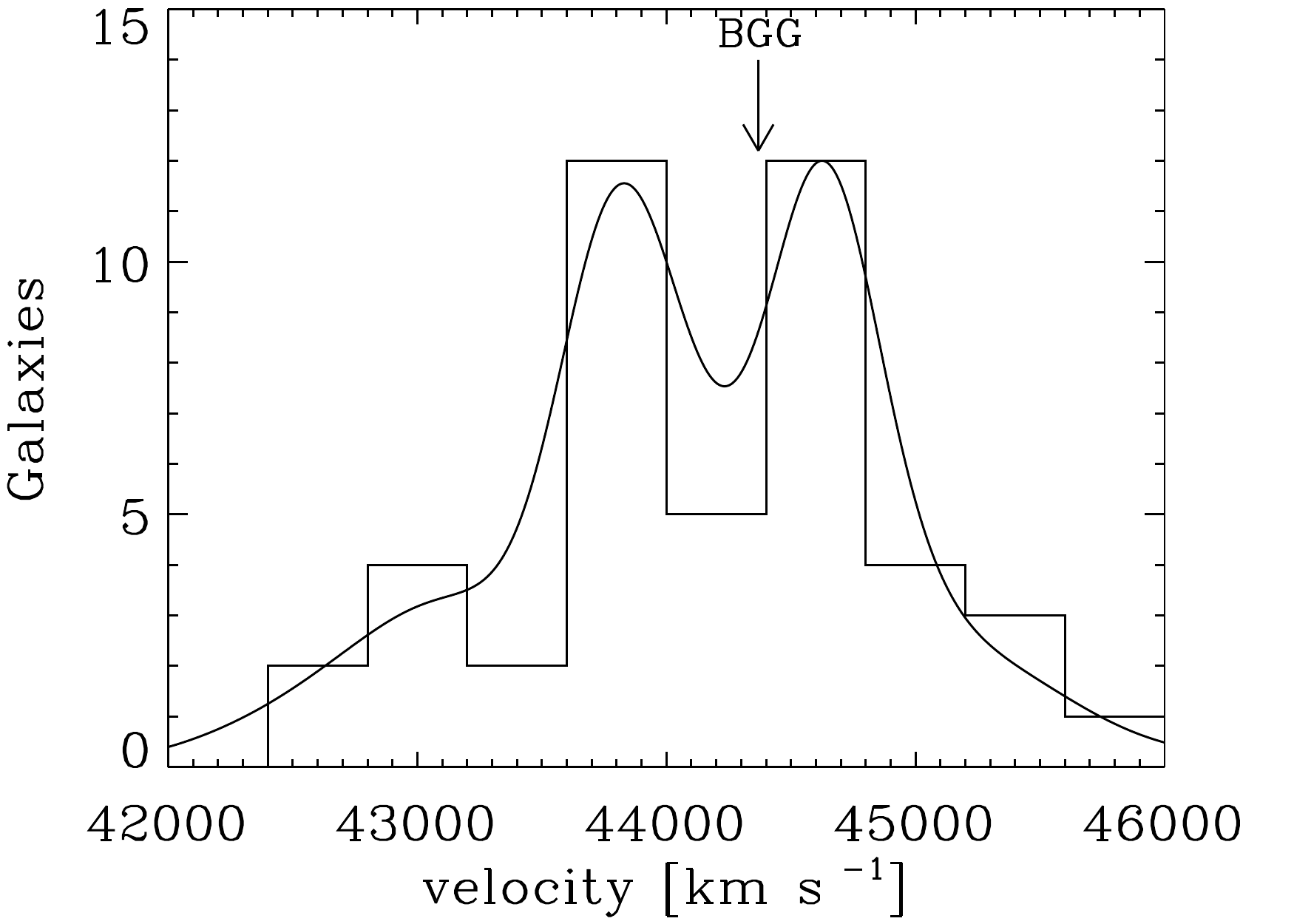} \\
\caption{The velocity histogram of FGS23. The curve represents the density reconstruction according to the 1D-DEDICA method. The vertical arrow indicates the velocity of the BGG.}
\label{vel_FGS23}
\end{figure}

In \citet{Zarattini2014} we presented the procedure we used for the galaxy member selection for our sample of galaxy systems. This procedure \citep[see][]{Fadda1996} is based on the 1D-DEDICA method \citep{Pisani1993}, that analyzes the density of galaxies in velocity space. For three FGs (FGS14, FGS23, and FGS26) the procedure detected a double peak in the velocity distribution near the velocity position of the BGG. However, the detected peaks are very close in velocity (rest frame differences $\le$ 1500 km s$^{-1}$, $\sim 2 \sigma_v$, where $ \sigma_v$ is the cluster's velocity dispersion) and with significant overlap in their galaxies. Thus we decided to consider them as single peaks in the member selection. Only in the case of FGS23 the significance of the two detected peaks in the velocity distribution is $>99\%$ c.l., thus supporting the presence of substructure. In this case, the peaks have 24 and 21 galaxies at $43\,385$ and $44\,623$ km s$^{-1}$, respectively. The peaks largely overlap (6/24 and 5/21 galaxies). In addition, FGS23 is the only FG for which such a bimodality also appears in re-applying the 1D-DEDICA procedure to the member galaxies alone. The velocity histogram of FGS23 is presented in Fig. \ref{vel_FGS23}.

\subsubsection{Peculiar velocity of the central galaxy}

BGGs are usually located in the center of the potential well of relaxed clusters \citep{Coziol2009}. Thus, another method to study the dynamical status of a system is to evaluate the velocity offsets of the BGGs with respect to the global recessional velocity of the cluster. For this estimation, we used the bootstrap test presented in detail in Sect. 4.2 of \citet{Gebhardt1991}. This test is proposed to be very rigorous and conservative compared to previous tests used in the literature. For the subsample of 5 FGs that we analyzed, only FGS14 seems to have a BGG with a peculiar velocity at a c.l. larger than 90\%. This value is considered as significant according to \citet{Gebhardt1991}. We present the rest-frame velocities of the central galaxies, as well as the mean velocity of each system together with its 1-$\sigma$ uncertainty in Table \ref{table:pec}.

\begin{table}
\centering
\caption{Rest-frame velocity of the BGG and mean velocity for the analyzed systems.}
\label{table:pec}
\begin{tabularx}{0.27\textwidth}{ccc}
\tiny
\\
\hline
\hline
name & v$_{BGG}$ & v$_{mean}$ \\
\multicolumn{1}{c}{} & km s$^{-1}$ & km s$^{-1}$  \\
\hline	
FGS02  & -172 & 69038 $\pm$ 188 \\
FGS14  & -348 & 66680 $\pm$ 124\\
FGS23  & -87   & 44150 $\pm$ 99\\
FGS27  & -83   & 55260 $\pm$ 113\\
FGS30  & 196  & 33919 $\pm$ 124 \\
\hline

\end{tabularx}
\end{table}

\subsection{Substructure in the projected position of galaxies}
In this section we study the presence of substructures by analyzing the projected positions of galaxies in the sky. We do not use kinematical properties, as explained in Sect. 2, so we apply this two-dimensional approach to the whole sample of 12 spectroscopically-confirmed FGs with $z \le 0.25$.

\subsubsection{2D-DEDICA}

\begin{figure}
\includegraphics[width=0.49\textwidth]{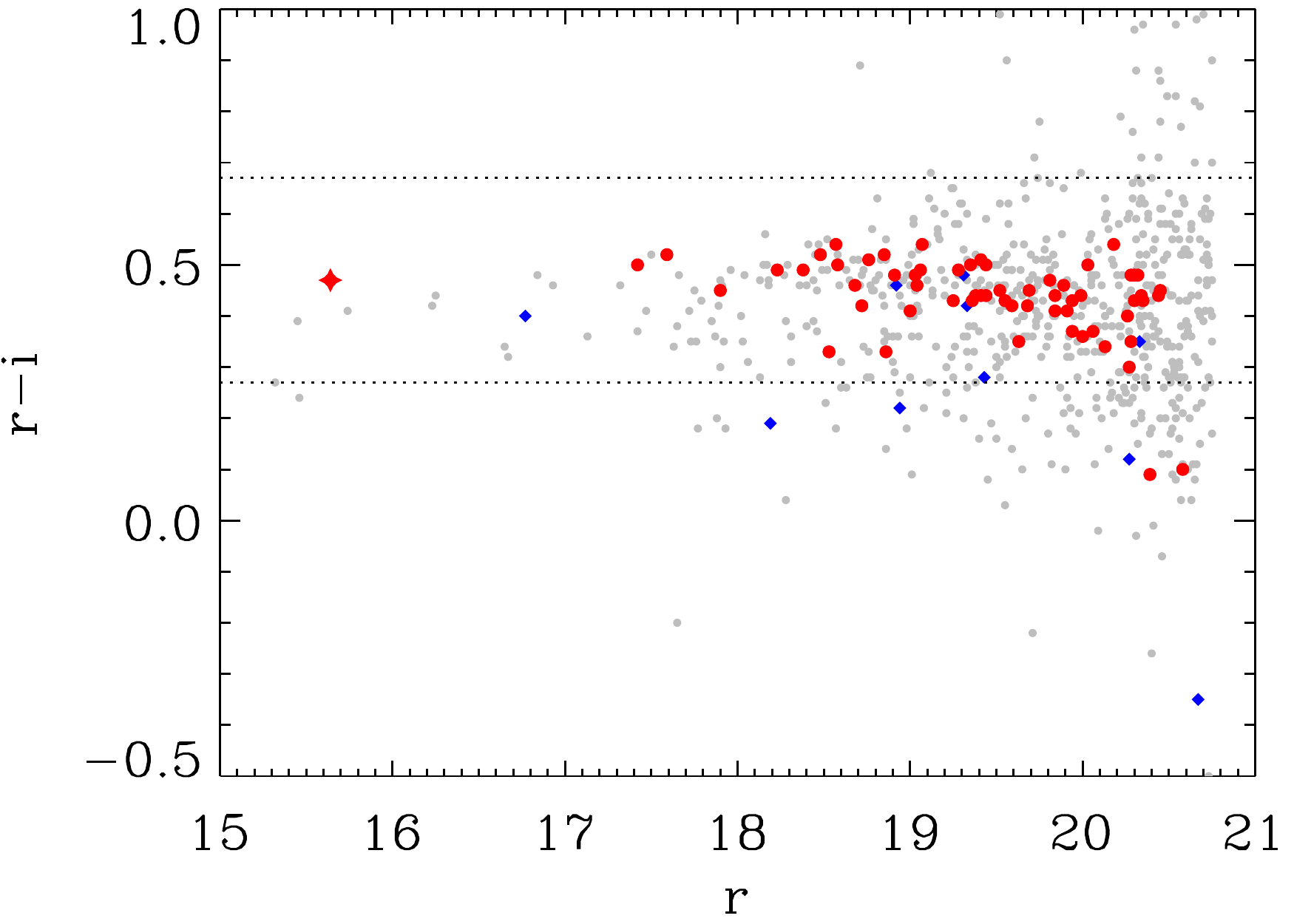} 
\caption{The $r$ vs $(r-i)$ diagram for FGS27. Grey circles represent all the galaxies within 2R$_{200}$. The red  cross is the BGG, whereas red circles and blue diamonds are spectroscopically-confirmed members and non-members, respectively. The horizontal dotted lines delimit the area in which we choose our ``likely members'' ($\pm 0.2$ from the color of the BGG). The color-magnitude diagrams for the other 11 systems in the sample can be found in the electronic version of the paper.}% Labels indicate the two peaks detected in the 2D analysis (see Table \ref{} and first panel of Fig. \ref{ded_vor}).}
\label{colmagFGS27}
\end{figure}

We were unable to study the presence of substructures using kinematical information in the outer regions of our FGs, since our spectroscopic data are mainly concentrated within 0.5 $R_{200}$, and, moreover, the velocity sample is not complete. In order to study the presence of substructure at larger radii, we analyzed the 2D distribution of galaxies brighter than $M_r=-19$ within 2 $R_{200}$. To minimize the effect of fore/background objects we only used galaxies located in a region of the color-magnitude space close to the position of the BGG, as already done in \citet{Girardi2014} . In particular, we defined the center of the red sequence as the horizontal line in the $r$ vs $(r-i)$ plane that crosses the BGG, considering the region with colors $\pm 0.2$ from that line as the red sequence \citep[see][]{Harrison2012} and the galaxies of this region as ``likely members''. As en example, we show the $r$ vs $(r-i)$ diagram for FGS27 in Fig. \ref{colmagFGS27}.

For our sample of 12 fossil systems we analyzed the distribution of galaxies within $R_{200}$ and 2 $R_{200}$, in order to detect false positives due to border effect. The 2D-DEDICA procedure \citep{Pisani1993, Pisani1996} looks for density peaks in the distribution of galaxies. To each identified peak, the procedure assigns a probability and a relative density $\rho_s$ with respect to the maximum-density peak. In Table \ref{tab:dedica2d} we summarize the peaks obtained by the 2D-DEDICA method within $R_{200}$, indicating also the number of galaxies and the position associated to each peak. We list all the peaks with significance $\ge 99\%$, with relative density $\rho_s \ge 0.5$, and with more than 10 galaxies.

In Fig. \ref{ded_vor} we show the contour maps obtained using the 2 $R_{200}$ analysis superimposing the peaks of the $R_{200}$ analysis. This was done to show the robustness of the results. In fact, the peaks coincide in almost all cases, with some exception near the $R_{200}$ radius due to border effects.

Four FGs present substructure according to the 2D-DEDICA analysis. They are FGS02, FGS14, FGS27, and FGS34. The results of the analysis of FGS02 and FGS14 showed clear signs of substructures. The former presented a subclump at NE position of the center, with 77 galaxies and a relative density $\rho_s=0.59$, the latter had a subclump at NE, with 33 galaxies and $\rho_s=0.5$. FGS27 had a second clump located SW to the center, with 33 galaxies and $\rho_s=0.5$. However, the peculiarity of this system was its central peak. In the $R_{200}$ analysis a single peak is detected, whereas in the 2 $R_{200}$ analysis the central region split into two peaks and the BGG was located between them. FGS34 seemed to be the more complex system. It had a clear subclump in the S region, with 28 galaxies and $\rho_s=0.71$, but it also had two other substructures. One was located NW from the center with 13 galaxies and $\rho_s=0.64$, the other was located N with 9 galaxies and $\rho_s=0.65$. These subclumps were close to our detection limit due to the small number of their galaxies. For this reason, to be conservative, we did not put them in Table \ref{tab:dedica2d}.

Finally, FGS29 and FGS32 were not clearly detected by the 2D-DEDICA procedure. In fact, the former appeared as a peak with $\rho_s=0.44$ with respect to the main peak of the area, that was located NW at a distance close to $R_{200}$. Probably, FGS29 is itself a subclump of a larger cluster. For FGS32, the procedure identified the main peak in a position that is $\sim$ 0.5 $R_{200}$ from the location of the BGG. A second peak with only 8 galaxies and $\rho_s=0.43$ is located E of the BGG. Moreover, only 30 galaxies are present in the region we analyzed and there is a larger cluster outside the virial radius. Therefore, also in this case the identification of the system associated to the BGG is not significant. For the other systems (FGS03, FGS17, FGS20, FGS23, FGS26, and FGS30) there is no substructure according to the 2D-DEDICA procedure. Nevertheless, although FGS30 does not present secondary peaks within $R_{200}$, it seems to stretch in the E-W direction. Moreover, the 2 $R_{200}$ analysis revealed another peak outside $R_{200}$ and located in the eastern region. This peak is aligned with the E-W stretching of FGS30.

\begin{table}
\caption[]{2D substructures detected in each fossil system of our sample. The position of the peaks refer to the $R_{200}$ analysis.}
\label{tab:dedica2d}
$$
\begin{array}{l c r c c c c }
\hline
\noalign{\smallskip}
\hline
\noalign{\smallskip}
{\rm name} & \mathrm{Subclump} & N_{\rm gal} & \alpha & \delta &\rho_{\rm S}\\
	 & & & ({\rm J}2000)&({\rm J}2000) & &\\%\arcmm:\arcs}&&\\
\hline
\noalign{\smallskip}
{\rm FGS02} &\mathrm{main}	 		& 54		& 01:52:41.4 	& +01:00:04	 &1.005\\
\vspace{0.2cm}
	 	 &\mathrm{NE}	 	& 77		& 01:52:46.1 	& +01:01:32	 &0.59\\
		 \vspace{0.2cm}
{\rm FGS03} ^\ast &\mathrm{main} 		& 36		& 07:52:16.9 	& +45:59:23	&1.00\\
{\rm FGS14} &\mathrm{main}			&144	& 11:46:49.5	& +09:52:40	&1.00\\
\vspace{0.2cm}
		 &\mathrm{NE}		& 33		& 11:47:00.2	& +09:55:20	&0.509\\
		 \vspace{0.2cm}
{\rm FGS17} &\mathrm{main} 		& 17		& 12:47:41.9	& +41:31:40	&1.00\\
\vspace{0.2cm}
{\rm FGS20} &\mathrm{main} 		& 36		& 14:10:04.4	& +41:45:42	&1.00\\
\vspace{0.2cm}
{\rm FGS23} &\mathrm{main} 		& 58		& 15:29:43.3	& +44:07:53	&1.00\\
\vspace{0.2cm}
{\rm FGS26} &\mathrm{main}			& 85		& 15:48:54.1	& +08:50:49	&1.00\\
{\rm FGS27}^\ast &\mathrm{main}			& 79		& 16:14:29.2	& +26:43:50	&1.00\\
\vspace{0.2cm}
		 &\mathrm{SW}		& 33		& 16:14:14.8	& +26:40:08	&0.50\\
		 \vspace{0.2cm}
{\rm (FGS29)}^\ast &\mathrm{main} 		& 19		& 16:46:58.2	& +38:49:39	&0.44\\
\vspace{0.2cm}
{\rm FGS30} &\mathrm{main} 		& 194	& 17:182:0.8	& +56:39:05	&1.00\\
\vspace{0.2cm}
{\rm (FGS32)} &\mathrm{main} 		& 22		& 17:28:41.6	& +55:15:22	&1.00\\
{\rm FGS34} &\mathrm{main} 		& 23		& 23:58:15.1	& +15:05:48	&1.00\\
		 & \mathrm{S}			& 28		& 23:58:10.5	& +15:03:21	&0.71\\
\noalign{\smallskip}
\noalign{\smallskip}
\hline
\noalign{\smallskip}
%\hline
\small
\end{array}
$$
\tiny Note: The 2D contours of the analysis within 2 $R_{200}$ are plotted in Fig. \ref{ded_vor}. Systems with $^\ast$ show minor differences in the peak position in the $R_{200}$ and 2 $R_{200}$ analysis.
\end{table}

\subsubsection{Voronoi tessellation and percolation test}
\label{Voronoi}
In order to check the robustness of our results, we applied a different procedure to the 12 FGs of our sample: the Voronoi Tessellation and Percolation (VTP) technique \citep[see][]{Ramella2001,Barrena2005}. The VTP is a non-parametric technique and does not require data smoothing. This means that the procedure does not assume any particular geometry during the structure detection process \citep[see][for a detailed explanation of the VTP technique]{Ebeling1993}. We applied the VTP out to 2 $R_{200}$ for each FG; this is necessary to have a better estimation of the background level, but (as done with 2D-DEDICA) we limited our analysis to the density peaks found within $R_{200}$. The results are shown in Fig.1, where blue and cyan circles (representing VTP peaks found at the 99.9\% and 99\% c.l., respectively) are superimposed to the 2D-DEDICA contour levels.

In agreement with the results of the 2D-DEDICA procedure, the VTP test identifies a double peak at the 99.9\% in FGS02 and FGS27. FGS14 and FGS34 (both showing the presence of substructure according to 2D-DEDICA) are detected as single density peaks by VTP (at the 99\% and only 96\%, respectively). Moreover, for FGS26 and FGS30, the VTP detects a central density peak at the 99.9\% and minor subclumps at the 99\% c.l., approximately following the elongation of the 2D-DEDICA contour levels. Finally, all the other FGs do not present any signature of substructures according to the VTP test.

\begin{figure*}
%\begin{table}
%\caption{FGS02 replicated 4 times}
\begin{tabularx}{\textwidth}{cc}
\tiny
\\
\centering
\includegraphics[width=0.42\textwidth]{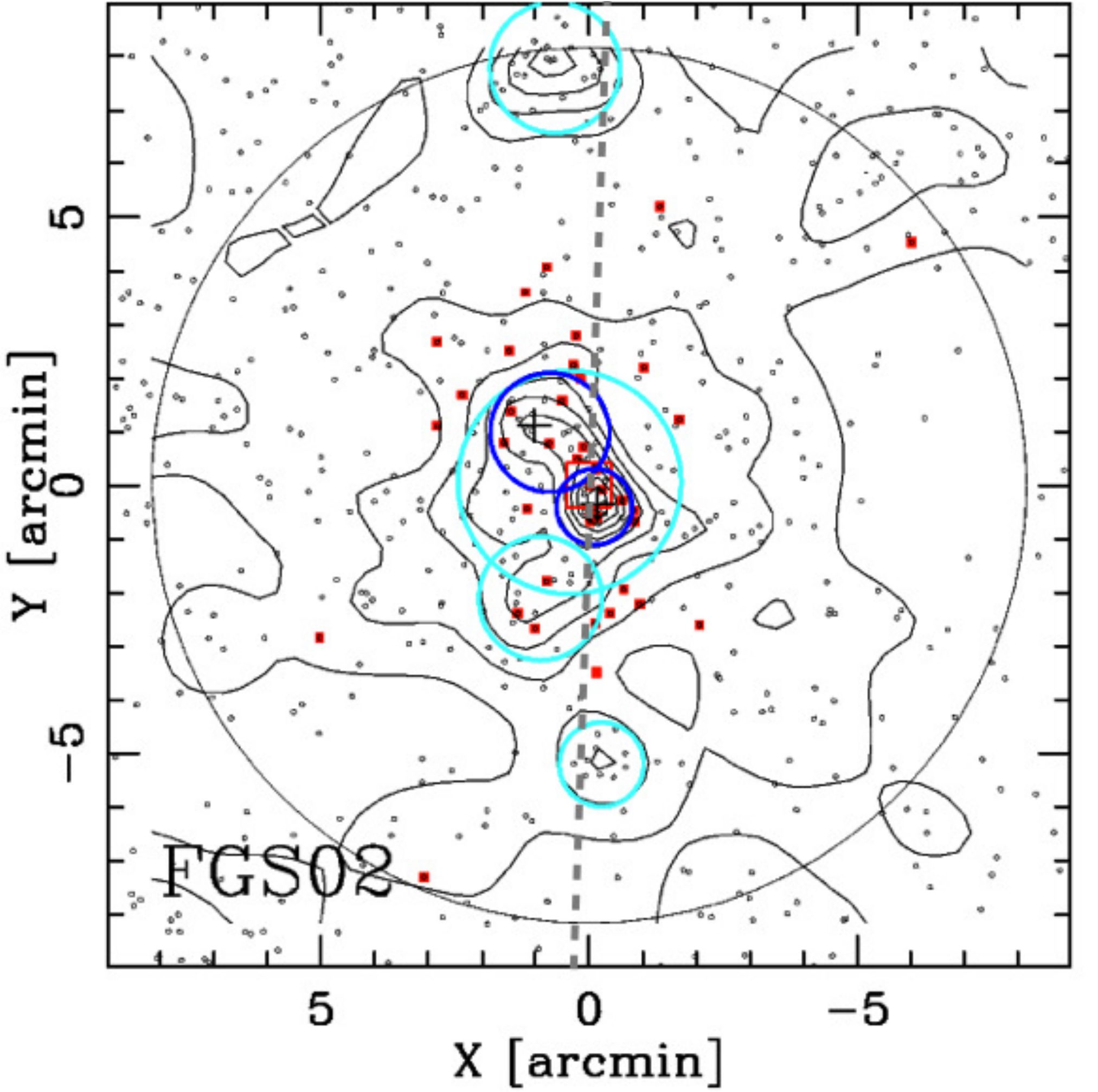} & \includegraphics[width=0.42\textwidth]{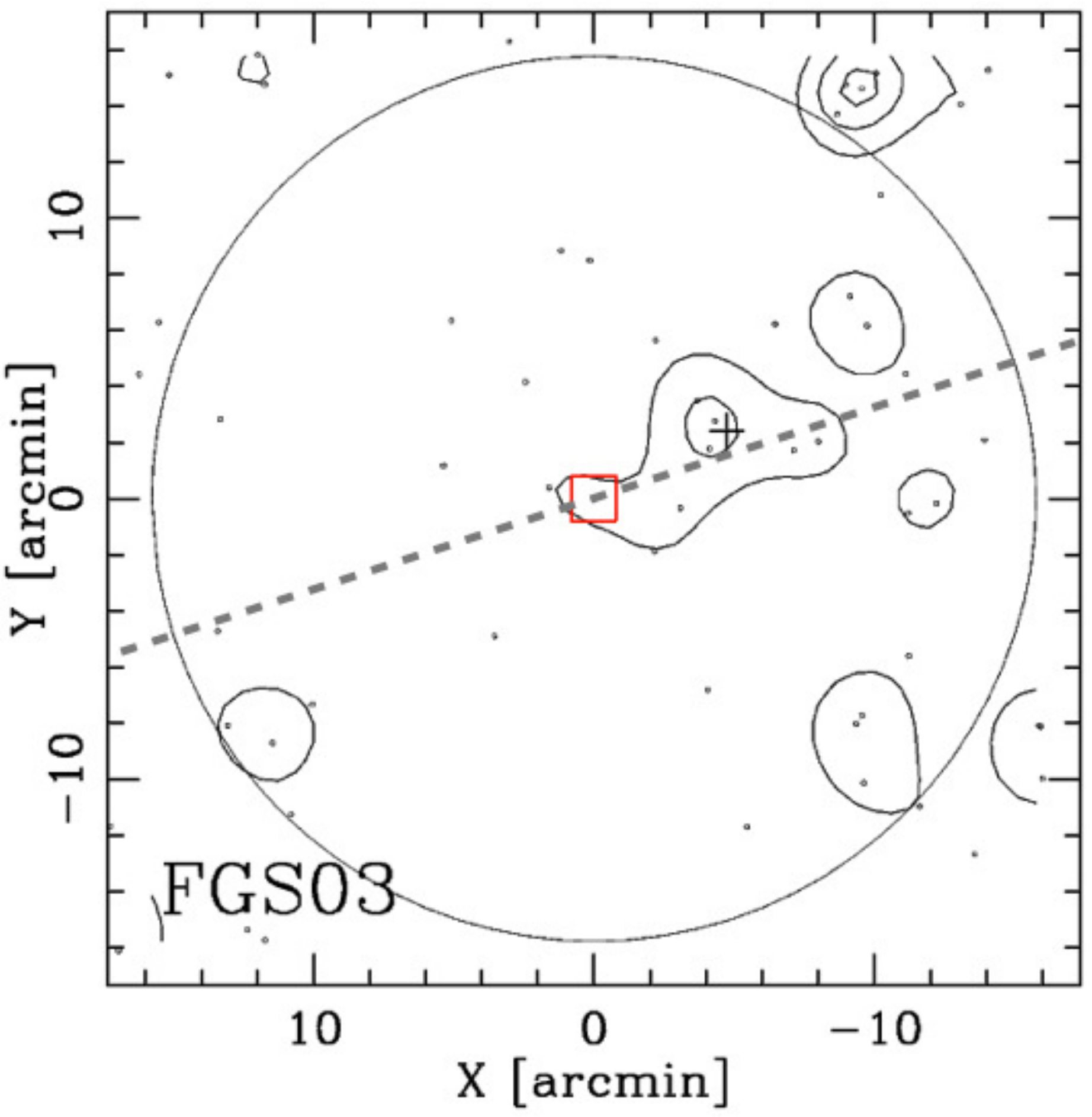} \\
\includegraphics[width=0.42\textwidth]{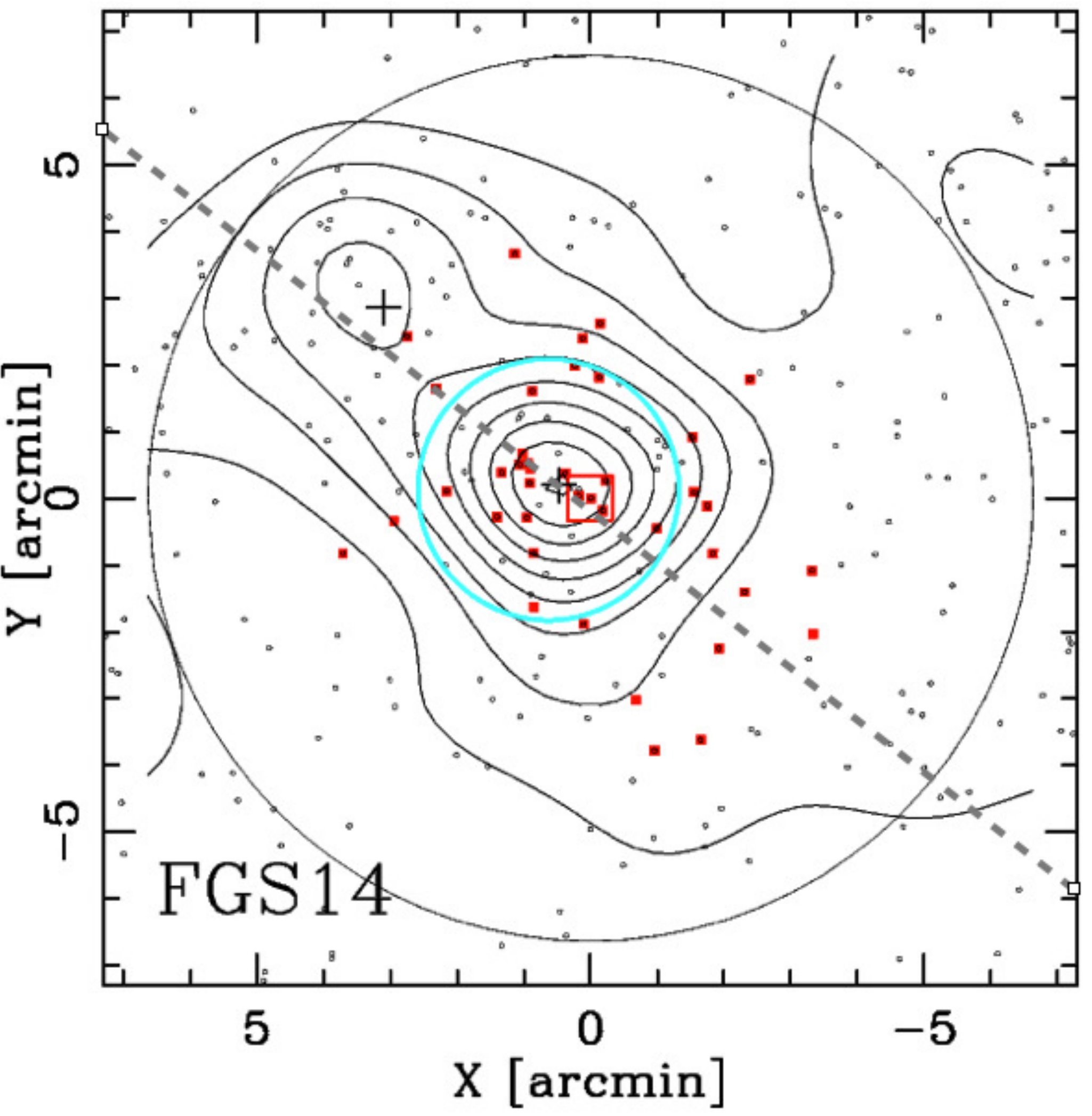} & \includegraphics[width=0.425\textwidth]{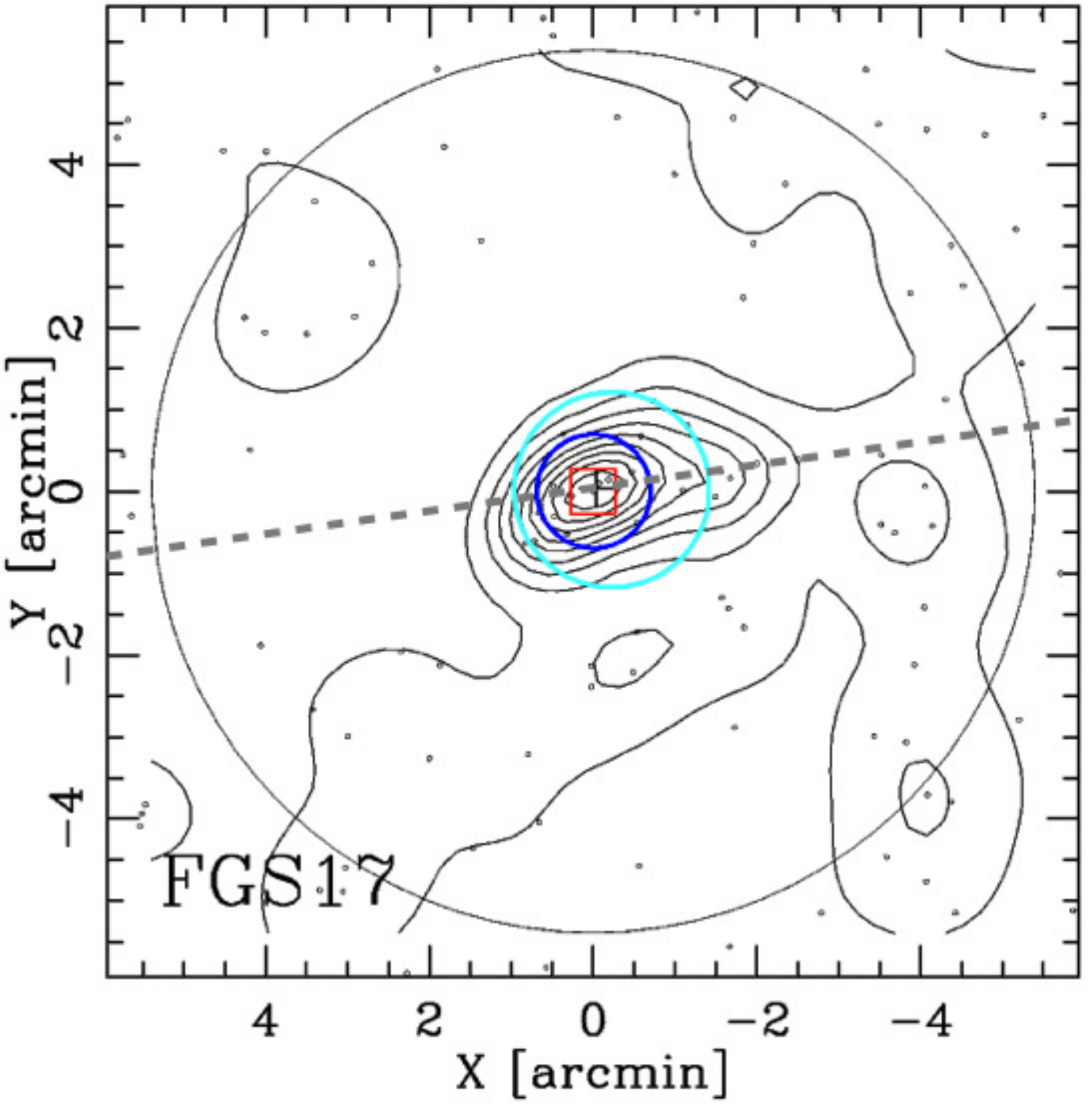} \\
\includegraphics[width=0.42\textwidth]{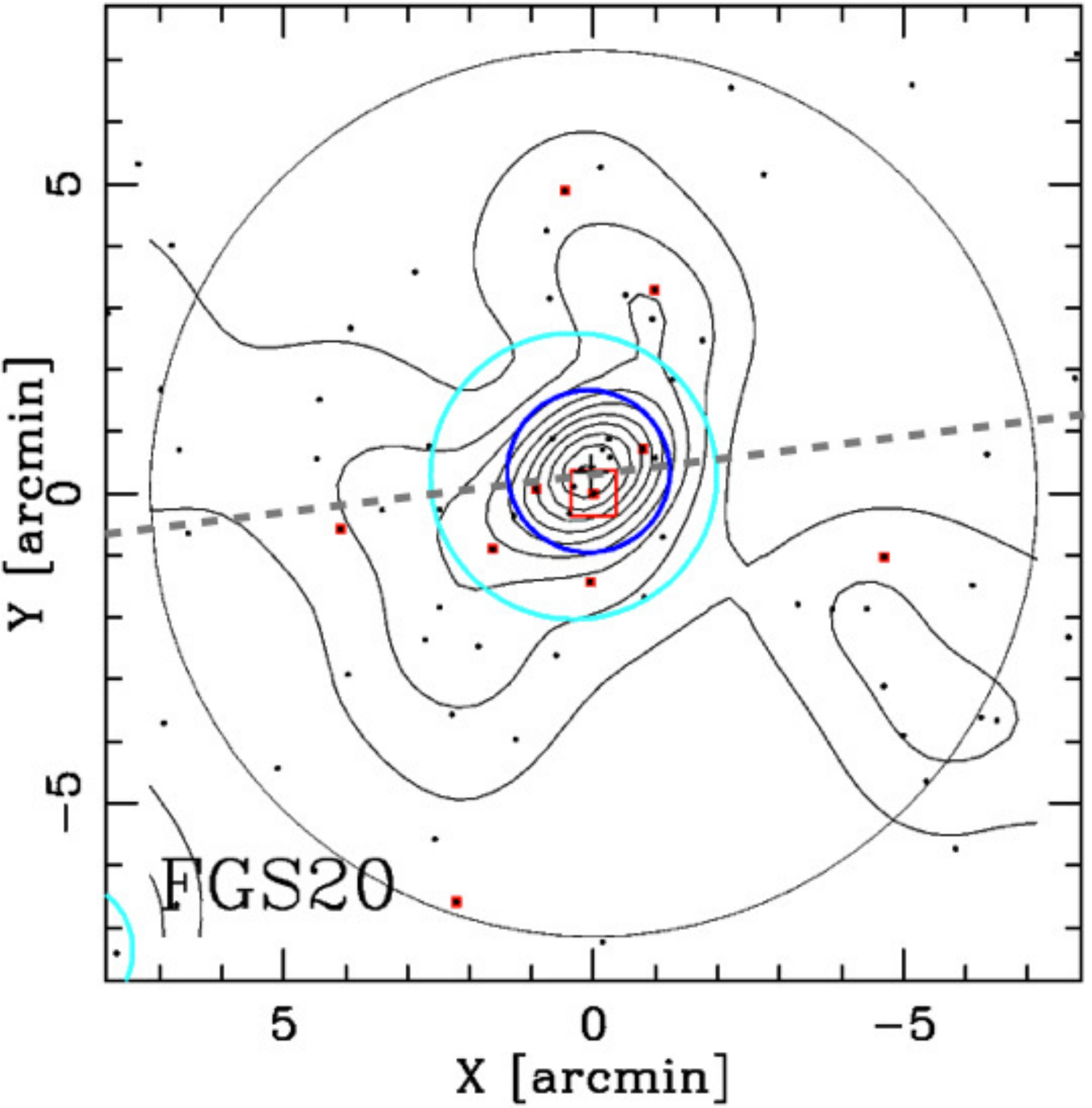} & \includegraphics[width=0.42\textwidth]{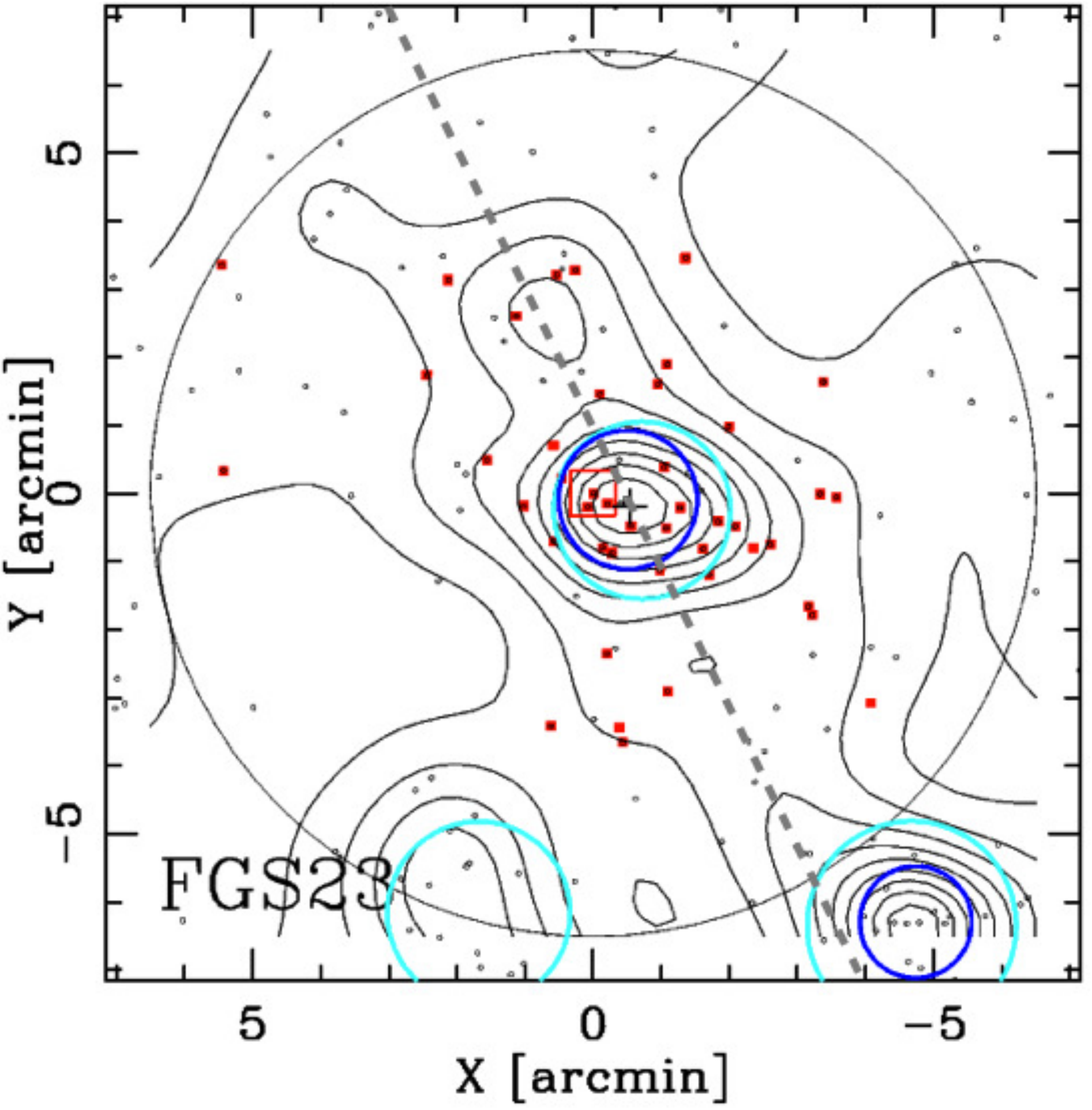} \\
\tiny
\\
\end{tabularx}
%\end{table}
\caption{2D-DEDICA contours (black) as obtained by the analysis of the substructures within 2 $R_{200}$ for the FGs in our sample. Black points are galaxies, red squares represent spectroscopically-confirmed members and, in particular, large open squares indicate the position of the BGGs. The black crosses are the centers of the significant peaks identified by 2D-DEDICA within $R_{200}$, which position is reported in Table \ref{tab:dedica2d}. Superimposed, blue circles are the peaks at 99.9\% c.l. and cyan circles are the peaks at 99\% c.l. as detected from the Voronoi procedure. The black circles are the $R_{200}$ radius. The grey dotted lines represent the PA of each system, as reported in Table \ref{ell}.}
\label{ded_vor}
\end{figure*}

\begin{figure*}
\ContinuedFloat
%\begin{table}
%\tabcolsep=0.2cm
%\caption{FGS02 replicated 4 times}
%\continuedfloat
\begin{tabularx}{\textwidth}{cc}
\tiny
\\
\centering
\includegraphics[width=0.42\textwidth]{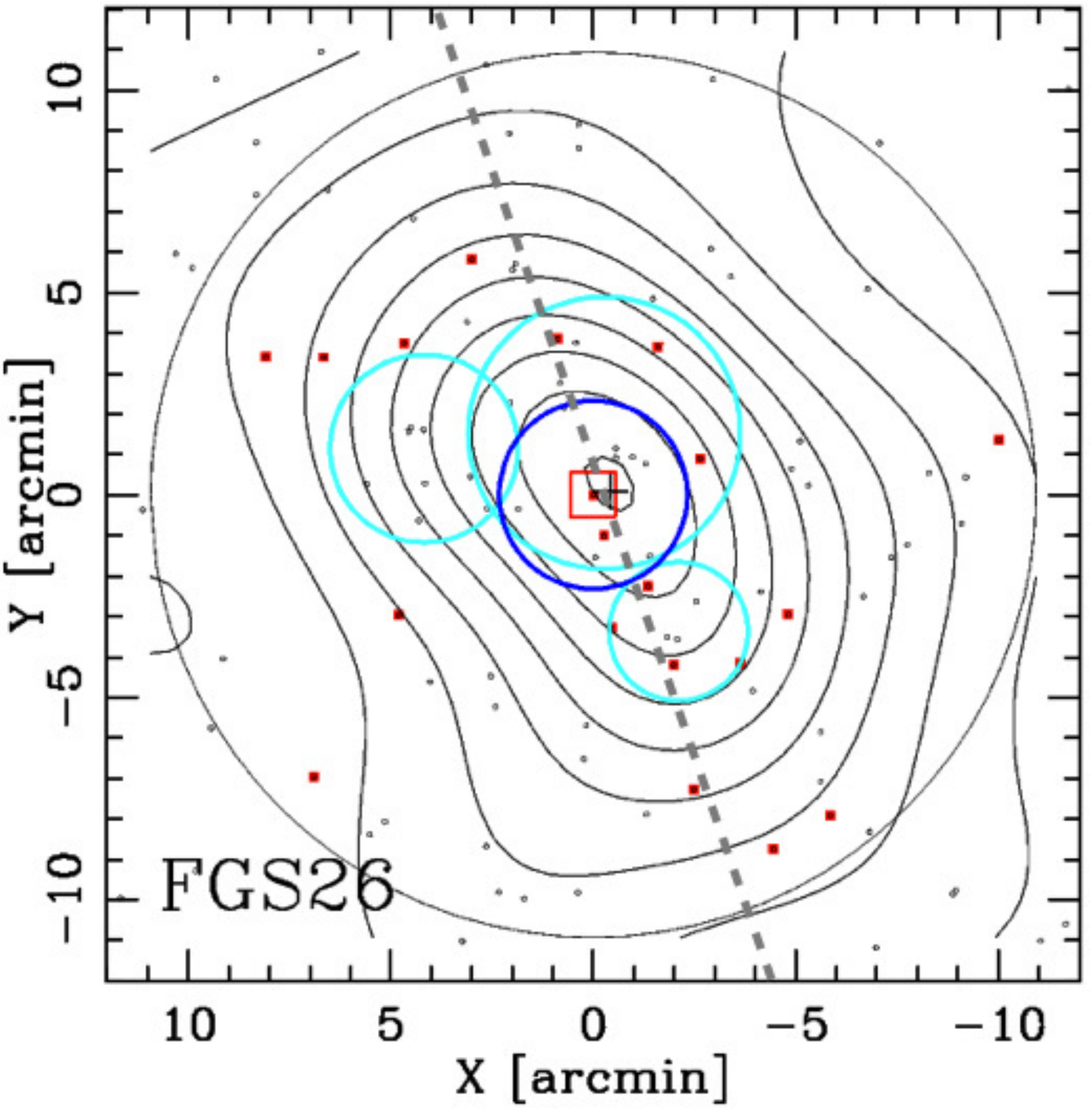} & \includegraphics[width=0.42\textwidth]{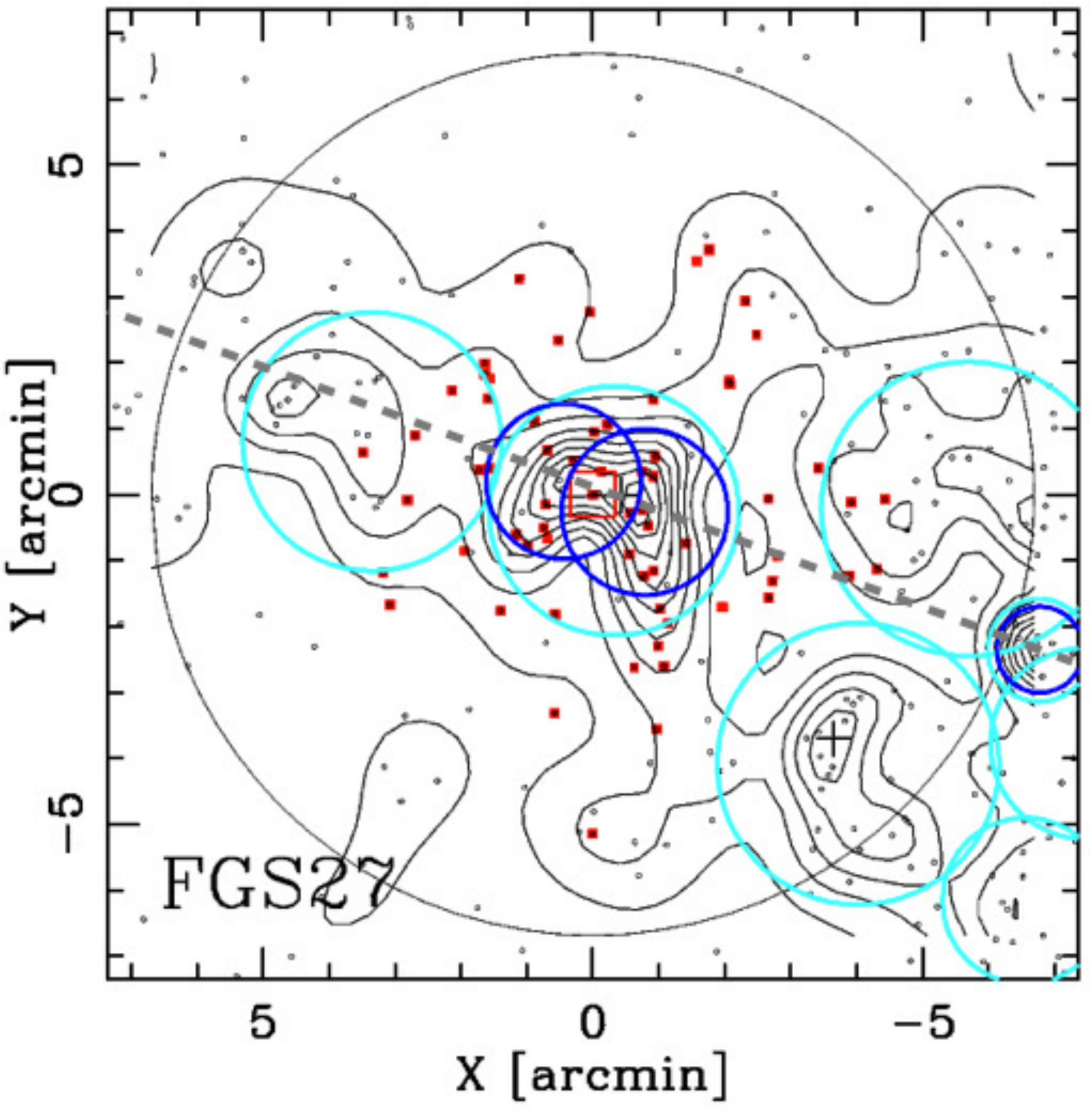} \\
\includegraphics[width=0.42\textwidth]{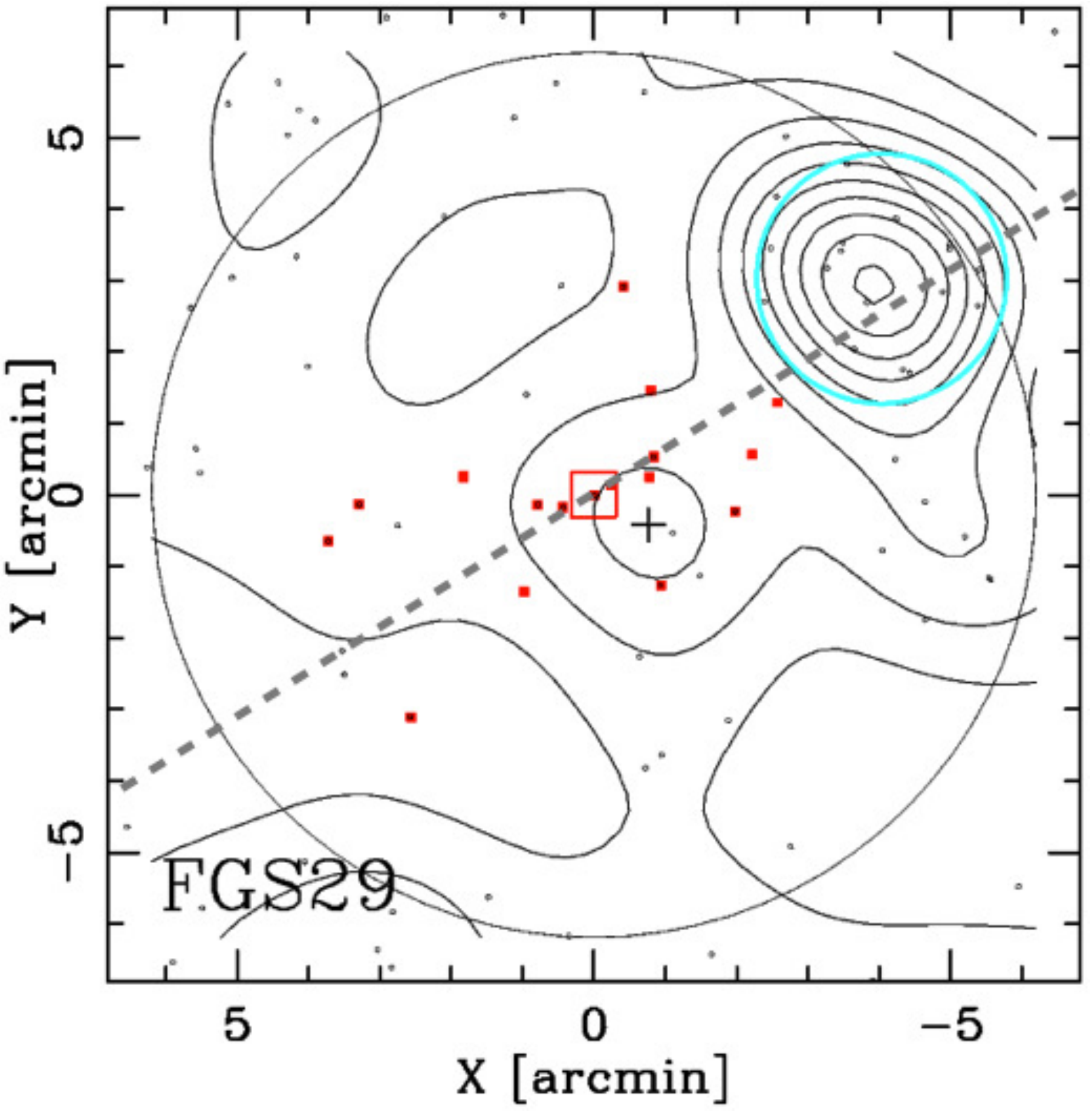} & \includegraphics[width=0.42\textwidth]{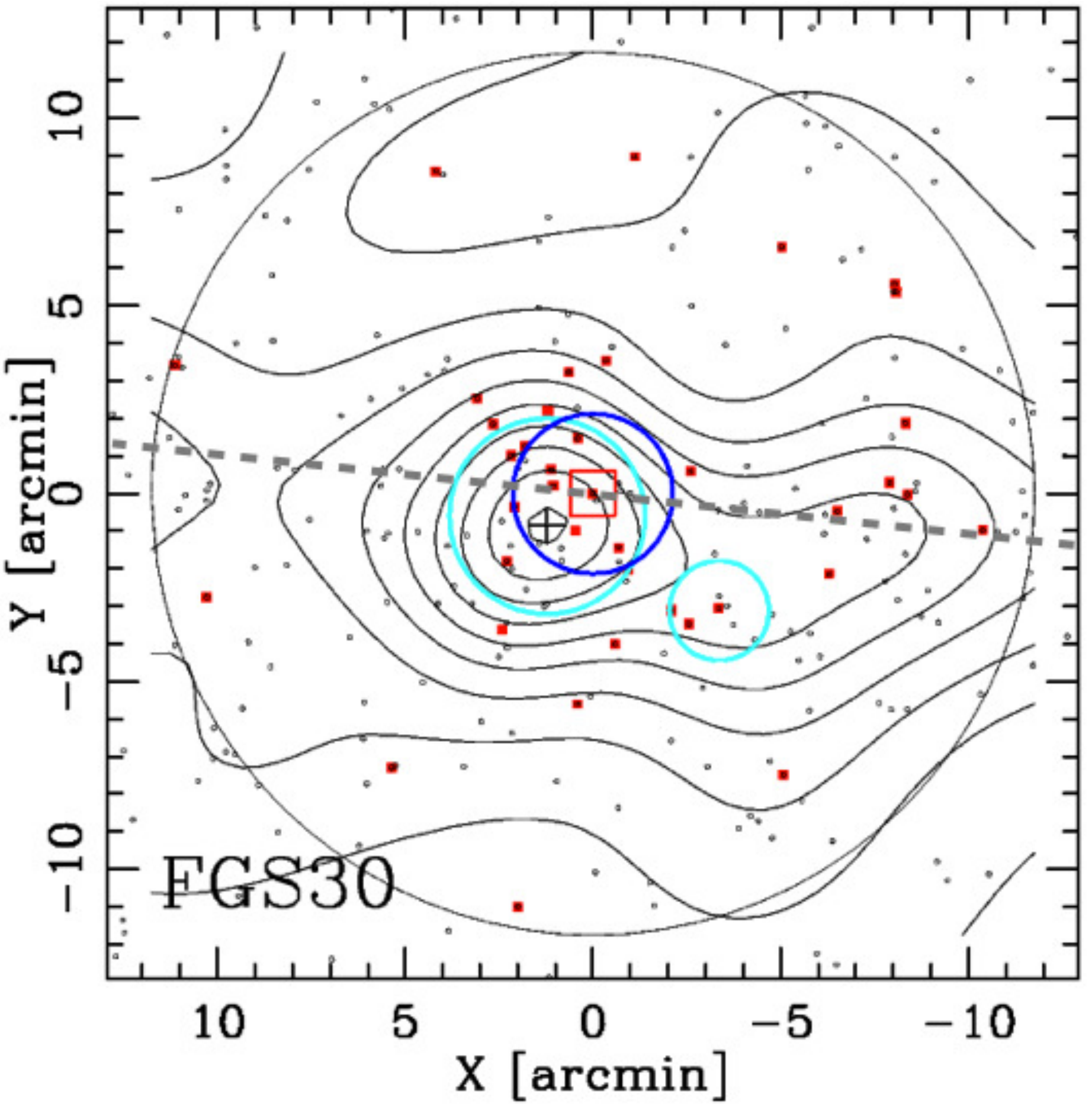} \\
\includegraphics[width=0.42\textwidth]{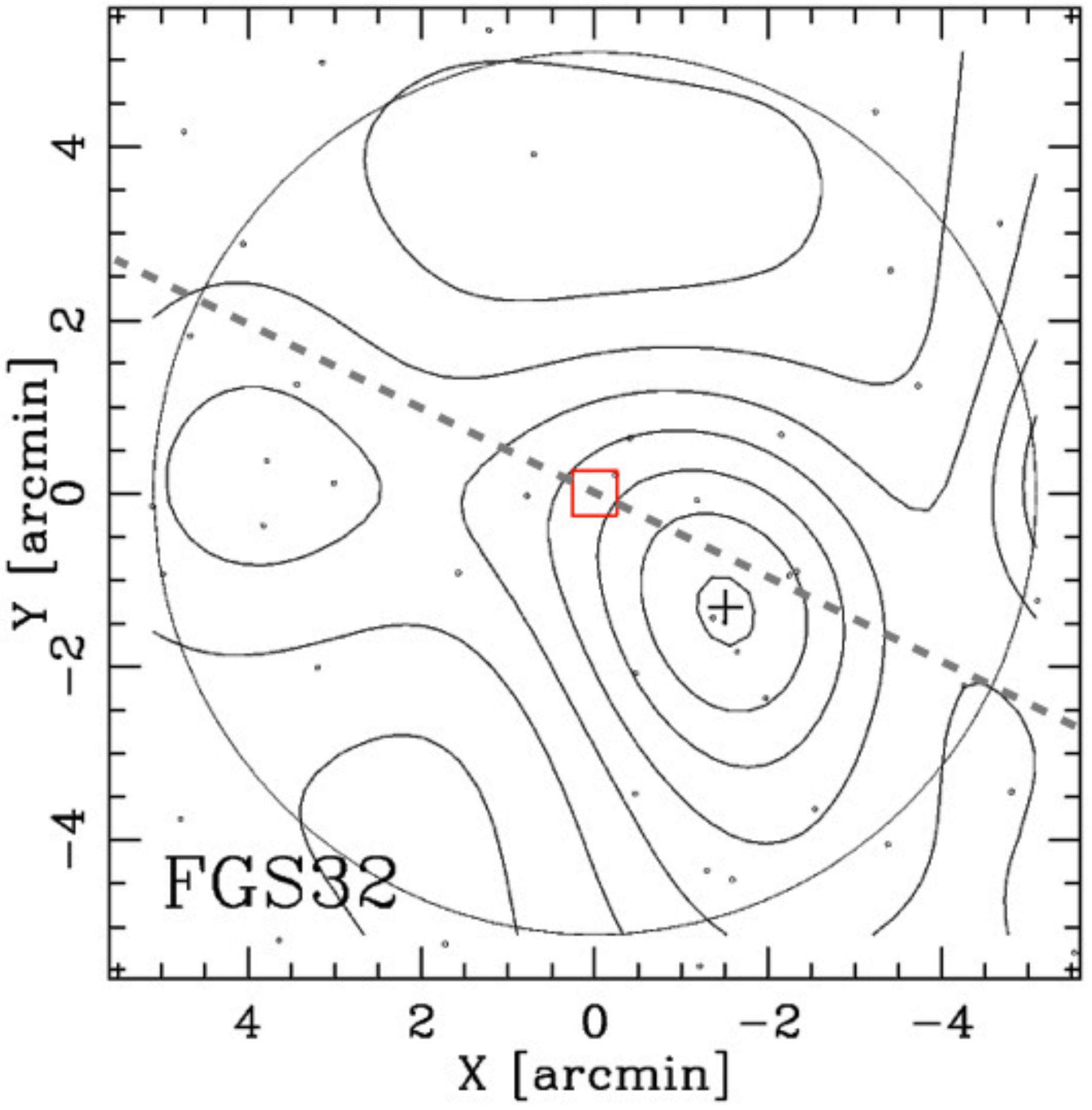} & \includegraphics[width=0.43\textwidth]{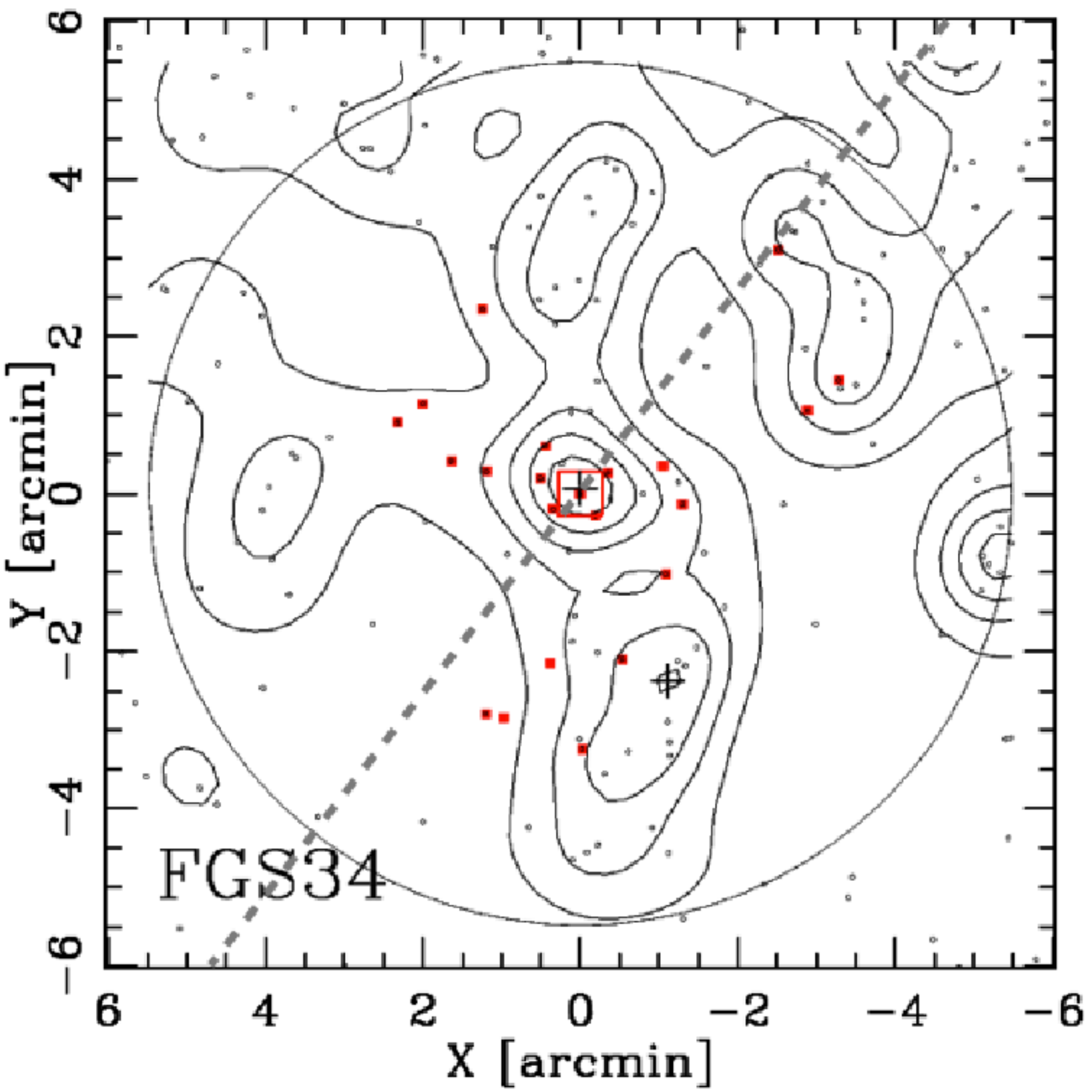} \\
\tiny
\\
\end{tabularx}
%\end{table}
\caption[]{(continued)}
\end{figure*}
\renewcommand{\thefigure}{\arabic{figure}}

\subsubsection{Ellipticity}
Relaxed systems are expected to be rounder than non-relaxed systems. In order to measure the shape of our FGs, we also computed the position angle of the major axis (PA) and ellipticity ($\epsilon$) of the galaxy distribution. Clusters tend to align with one another, mainly due to the anisotropic merging induced by the presence of the large-scale structure filaments \citep{Basilakos2006}. We followed the approach of \citet{Carter1980}. A detailed description of the method can be found in \citet{Basilakos2000} and, in particular, we used the procedure described in their discrete case. The method is based on the computation of the dispersion ellipse, meaning the contour at which the density is 0.61 times the maximum density of the galaxies in the cluster. The contour was determined using the first five moments of the observed distribution. From these moments, we obtained the center, PA, and $\epsilon$ of the distribution. In Table \ref{ell} we list the PA and $\epsilon$ for our FG sample. Only FGS27 and FGS30 are not circular to 3-$\sigma$ c.l.. In both cases, the maps we presented in Fig. \ref{ded_vor} show an elongation that is oriented in a direction which is consistent with the PA we detected.

\subsection{Substructure in the velocity-position field}
\label{3D}
\begin{figure*}
\begin{tabularx}{\textwidth}{ll}
%\numberwithin{figure}{chapter}
\includegraphics[width=0.48\textwidth]{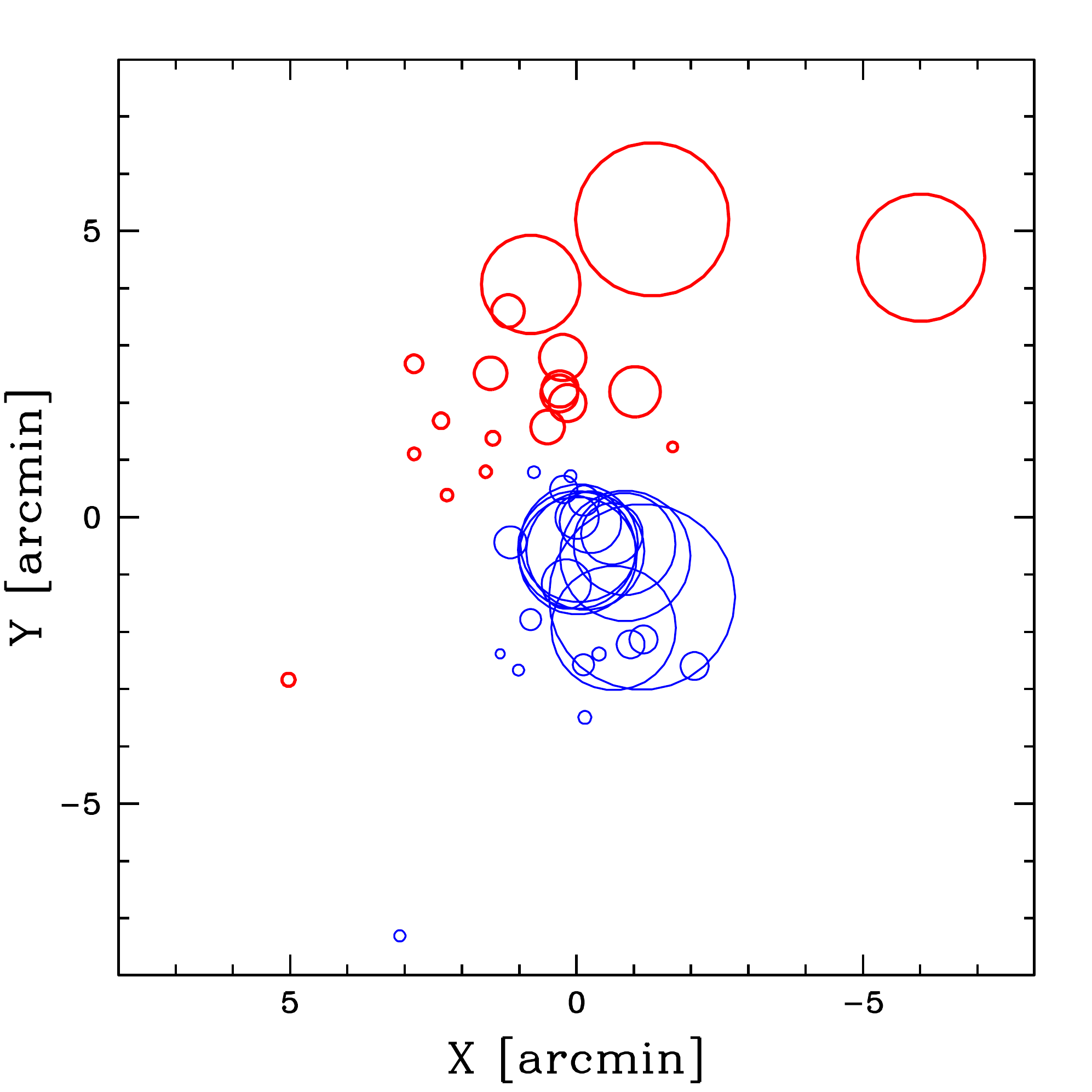} & \includegraphics[width=0.48\textwidth]{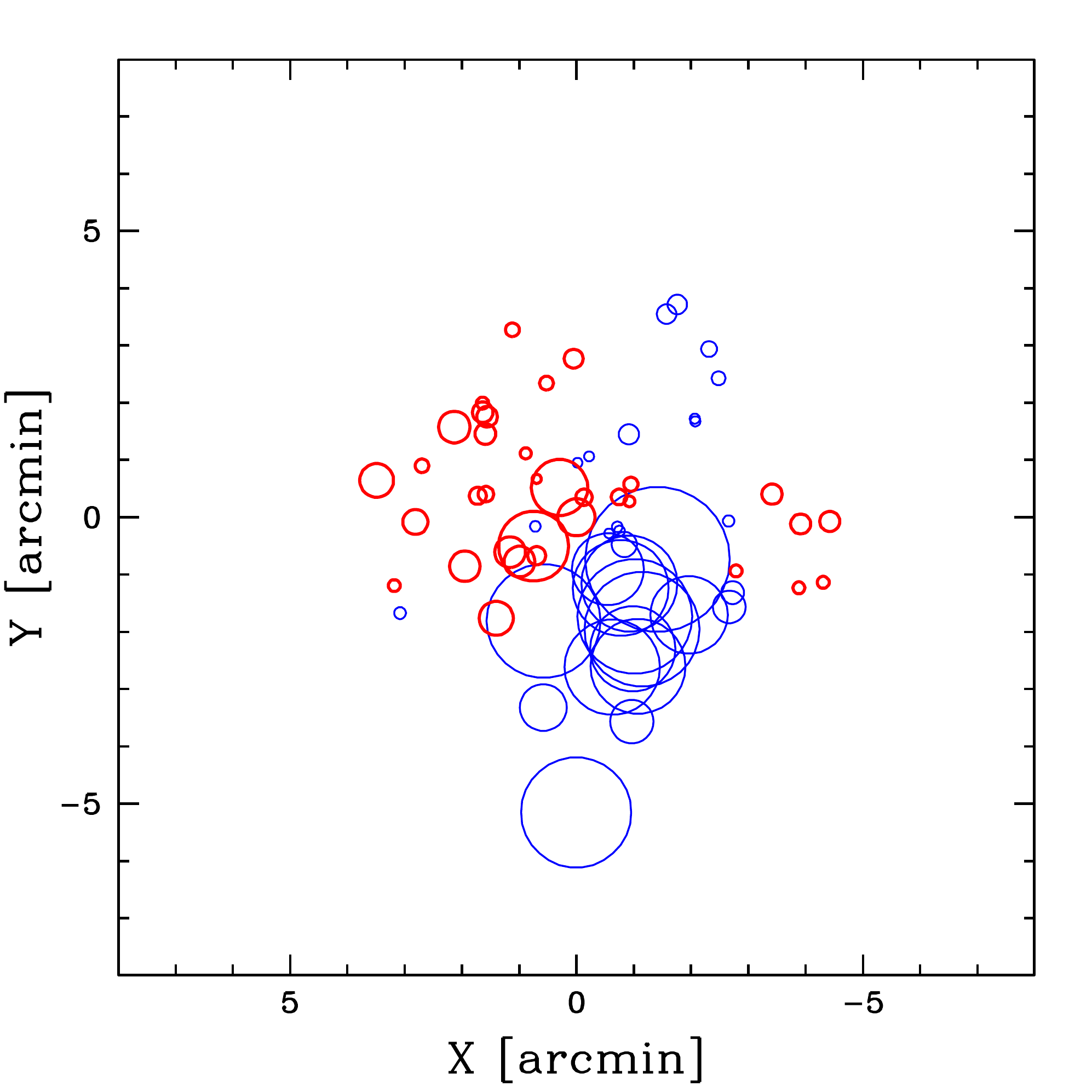} \\
\end{tabularx}
\caption{DS bubble plot for FGS02 (left panel) and FGS27 (right panel). The larger the circle, the larger the deviation of the local mean velocity from the global mean velocity. Blue and heavy red circles show where the local value of mean velocity is smaller or larger than the global value, respectively.}% Labels indicate the two peaks detected in the 2D analysis (see Table \ref{} and first panel of Fig. \ref{ded_vor}).}
\label{DS}
\end{figure*}

In general, the existence of correlations between positions and velocities of cluster galaxies is explained by the presence of substructures. To investigate this aspect, we combined both velocity and position information to compute the $\Delta$-statistics proposed by \citet{Dressler1988}. The DS test is capable to detect spatially compact subsystems whose velocity differs from the average velocity of the system or whose velocity dispersion differs from the global one, or both. For the $i$-th galaxy, the method computes its $\delta_i$ parameter, which is defined as
\begin{equation}
\delta_i^2 = \left(\frac{N_{\rm{nn}}+1}{\sigma_{v}^2}\right)\,\Bigg[\,\bigg(\left<v\right>_{\rm loc} -\left<v\right>\bigg)^2+\bigg(\sigma_{v,{\rm loc}} - \sigma_{v}\bigg)^2\,\Bigg].
\end{equation}
The subscript loc refers to local quantities, computed over the $N_{\rm{nn}}=10$ closest neighbors of the $i$-th galaxy, whereas the $\left<v\right>$ and
$\sigma_v$ are the global values of the mean velocity and velocity dispersion of the cluster, respectively. Once all the individual $\delta_i$ were derived, the $\Delta$-statistic was computed as the sum of all the $N$ individuals. Then, the procedure performed 1000 Monte Carlo (MC) simulations by randomly shuffling the velocities within galaxies, maintaining constant their positions. For each simulation, the parameter $\Delta$ was computed. We obtained a distribution of $\Delta$ over the 1000 MC simulations and determined the significance of the $\Delta$ estimated on observed galaxies. The DS test needs a statistically significant number of galaxies to be reliable \citep{Pinkney1996}. For this reason, we decided to apply it only to those systems with at least 30 confirmed members (see Sect. \ref{1D}). These systems are FGS02, FGS14, FGS23, FGS27, and FGS30. FGS02 shows substructure at $\sim 99\%$ c.l., FGS27 at $\sim 95\%$ c.l., whereas the substructure in the other three FGs is not significant. We show the DS bubble plot for FGS02 and FGS27 in Fig. \ref{DS}.

Moreover, we used two other estimators that are alternative to $\delta_i$ and analyzed only the differences in velocity ($\delta_{i,v}$) and in velocity dispersion ($\delta_{i,\sigma}$). The definition of these quantities follows the prescription of \citet{Girardi1997} and \citet{Barrena2011}
\begin{equation}
\label{DS_v}
\delta_{i,v}= \bigg[\frac{(N_{\rm nn}+1)^{1/2}}{\sigma_v}\bigg]\times \bigg(\left<v\right>_{\rm loc} -\left<v\right> \bigg),
\end{equation}
\begin{equation}
\label{DS_sigma}
\delta_{i,{\sigma}}= \bigg[\frac{(N_{\rm nn}+1)^{1/2}}{\sigma_v}\bigg]\times \bigg(\sigma_{v,{\rm loc}}- \sigma_v \bigg).
\end{equation}
Only FGS02 and FGS27 were found to show some hint of substructures. In particular, FGS02 presents a substructure in the velocity indicator at $>99\%$ c.l., whereas FGS27 shows the signature of substructure in both indicators at $\sim 92\%$ c.l..

\subsubsection{Presence of a velocity gradient}
The cluster velocity field can be affected by the presence of other structures, such as nearby clusters, superclusters, or filaments. If these structures are asymmetric with respect to the cluster, they will give rise to a velocity gradient in the cluster velocity field. We looked for the presence of a velocity gradient in our sample of FGs by performing a linear least-squares fit to the observed velocities with respect to the galaxy position in the plane of the sky, i.e. we fit a plane surface to the data in the (RA, Dec, $v_{LOS}$)-space. For each system, we computed both the velocity gradient and the coefficient of multiple determination $R^2$, which measure the deviation of the dependent variable owing to the set of the two independent variables. Thus, we performed 1000 Monte Carlo simulations that shuffle the velocities while keeping the positions fixed and we computed the $R^2$ coefficient every time. The significance of the gradient was defined as the fraction of times in which the $R^2$ we obtained from simulations was smaller that the observed one. The highest significance turns out to be that of FGS02 (89\% c.l.), followed by FGS23 (57\%), FGS14 (38\%), FGS30 (13\%), and FGS27 (18\%). Thus, according to this test, none of the fossil systems that we analyzed showed signs of substructure.

\subsection{Interactions in the region of the central galaxy}
\label{BGGsub}
We looked for signs of strong ongoing interactions around the central galaxies of our FG sample. In the merging scenario, FGs formed at high redshift and had enough time to merge all the $M^\ast$ galaxies in the central object. Then, they evolved via minor mergers only. However, the last major merger for the BGG could have occurred at a more recent epoch in FGs than in non-FGs \citep{Diaz-Gimenez2008}. 

To investigate this aspect, we used our own deep $r$-band images for 10 systems, and SDSS images for an additional two systems (FGS26, FGS27) \citep[see][for details]{Zarattini2014}. For each FG, we used the {\tt ellipse} package of IRAF to fit elliptical isophotes to the two-dimensional image of the BGG. Successively, we adopted the IRAF task {\tt bmodel} to generate the photometric model of our galaxy. Finally, we subtracted the model from the original image, as done in \citet{Zarattini2014}, and analyzed the residuals. 

We found that none of the inspected galaxies from our sample of 12 spectroscopically-confirmed FGs show signs of ongoing interactions. In particular, FGS29 seems to have an interaction in the W side, but there is no clear tail. Moreover, FGS34 seems to have two tails in the SE side. These tails are probably far objects or gravitational arcs because they are not connected to any galaxy. All the other systems show no sign of ongoing interaction.

%\subsubsection{FGS28}

\begin{figure}
\centering%\begin{table}
%\caption{FGS02 replicated 4 times}
\includegraphics[angle=90, width=0.45\textwidth]{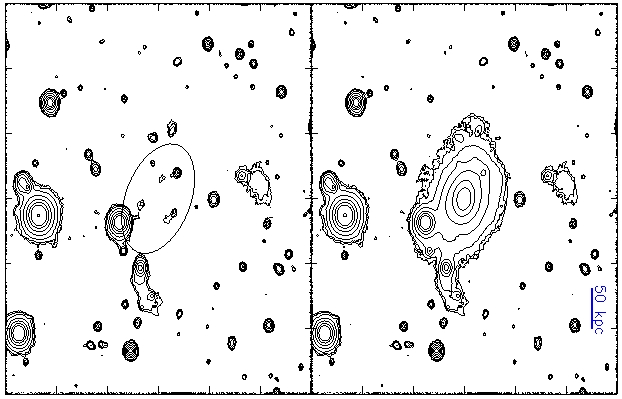} \\
\caption{Constant SB profiles and residuals for FGS28. The contours represents SB from 21 to 27 mag arcs$^{-2}$, separated by 1 mag arcsecs$^{-2}$. In the contour images, the blue line is equivalent to 50 kpc at the distance of the FGS28. In the residual images, the ellipse represent the 25 mag arcs$^{-2}$ of the BGGs (see text for details).}
\label{BGG1}
\end{figure}

We were able to analyze the presence of ongoing interactions also for FGS28. This system is the smallest and poorest of the \citet{Santos2007} sample, and its spectroscopic and photometric catalogs are so poor that we were not able to apply any other test to it. For this reason, it is not included in our sample of 12 FGs or in Table \ref{summary}. However, we found that FGS28 has two evident tails on the W and NNE side (see Fig. \ref{BGG1}). These tails are connected to two galaxies that seem to be involved in a merging process with the BGG. This is the first evidence of ongoing interactions around the BGG of low redshift FGs to date (see \citealt{Ulmer2005} and \citealt{Irwin2015} for two higher-redshift examples). The only similar result is that presented in \citet{Lieder2013}, who found a galaxy being disrupted in the NGC 6482 FG. In their case, the galaxy that is disrupting the smaller one is not the BGG, it is instead the third brightest member of the system, located at about 0.75 $R_{200}$. 

%%%%%%%%%%%%%%%%%%%%%%%%%%%%%%%%%%
%%%%%%%%%%%%% SECTION 4%%% %%%%%%%%%%%%
%%%%%%%%%%%%%%%%%%%%%%%%%%%%%%%%%%
\section{Discussion}
\label{discussion}

\subsection{Caveats of the statistical analysis}

As shown in Sect. \ref{results}, there is not a single statistical test that resolves the problem of detecting substructures in galaxy clusters. Some of the tests are more sensitive to particular configurations of substructure than others. Thus, the best approach is to use several statistical tests for the identification of the substructure.
Following the complete analysis of \citet{Pinkney1996}, the 1D tests are in general less sensitive with respect to 2D and 3D tests. Nevertheless, 1D tests are more effective in detecting substructures that are superimposed to the main body of the system, and rapidly moving in the line-of-sight direction. Similarly, in the case of a recent merger of two unequal-sized systems, the asymmetries in the velocity field can correctly detect the substructure.

The 2D tests work in the coordinates space. In our case, we worked with photometric members only, without any spectroscopic information. However, we used red-sequence galaxies, that are thought to be the best tracers of substructures \citep{Lubin2000}.

\citet{Pinkney1996} found that the DS test is the most sensitive to the presence of substructures but, at the same time, it has the highest detection of false positives due to the elongation and velocity dispersion of clusters. Moreover, the DS test has two well-known weaknesses. It is not able to detect subclumps when they are superimposed and when they have the same $\sigma_v$ and mean velocity. In addition, \citet{Cohn2012} showed that the DS test tends to detect substructures more efficiently in systems with larger subclumps.

\subsection{Comparison with other samples}

The 2D procedures we adopted (2D-DEDICA and VTP) showed that BGGs in FGs are located in dynamically different environments. We found that 4 FGs analyzed with the 2D-DEDICA (FGS02, FGS14, FGS27, and FGS34) and 4 FGs studied with the VTP method (FGS02, FGS26, FG27, and FGS30) show significant substructures. This corresponds to a fraction of $\sim$ 30\%. Comparison with the literature is not straightforward. In fact, the results depend on different details, such as the photometric limits, radius, number of members, and methods adopted. For example, \citet{Geller1982} analyzed a sample of 65 rich clusters using two-dimensional contours. They found that $\sim$ 40\% of their clusters presented significant substructures. On the contrary, \citet{West1988} analyzed 55 clusters, finding that the fraction of systems with substructures is compatible with the statistical fluctuations. Also \citet{Ramella2007} studied the fraction of clusters with substructures in the WINGS sample \citep{Fasano2006} using the 2D-DEDICA procedure. They found substructures in 71\% of their clusters. Nevertheless, they used a much larger sample of galaxies (from $3\,000$ to $10\,000$ per cluster). \citet{Ramella2007} noted that the quality of the photometric catalog is a key point in the detection of substructures, as well as the adopted algorithms. In a recent work, \citet{Wen2013} have developed another method for the study of the relaxation state of 2092 clusters. They have estimated that $\sim$ 70\% of the sample clusters have substructures. Only 20 systems show a magnitude gap larger than 2 and they are found to be relaxed.
According to our results, the fraction of systems with substructures is similar in FGs and non-FGs. However, the small number of systems we analyzed and the apparent discrepancy with \citet{Wen2013} suggest that a larger sample is needed.

The analysis of the ellipticity of 12 FGs shows that they display a modest elongation. Only FGS27 and FGS30 are remarkably elongated. This is in agreement with the analysis of 25 clusters performed by \citet{DeFilippis2005}, in which they claimed that none of their clusters show an extreme ellipticity. The median value of the ellipticity for our sample of 12 FGs is $\langle \epsilon \rangle=0.18\pm0.09$. \citet{Binggeli1982} studied a sample of 44 Abell clusters and measured a median value $\langle \epsilon \rangle=0.25\pm0.12$. \citet{deTheije1995} found $\langle \epsilon \rangle=0.4$ for 99 low-redshift Abell clusters, but they corrected the ellipticity using simulations. Indeed, they claimed that the value of $\epsilon$ they presented in the paper is, on average, higher than what it is found without applying any correction. Therefore, although our results are not in contradiction with previous works, a direct comparison is not possible in this case. More recent works \citep[e.g.,][]{DeFilippis2005,Lee2006} are once again of difficult comparison due to the wide variety of techniques adopted to compute the ellipticity.

\begin{table}
\centering
\caption{Position angle and ellipticity of the galaxy distribution in fossil systems}
\label{ell}
\begin{tabularx}{0.27\textwidth}{lcccc}
\tiny
\\
\hline
\hline
name & N$_{\rm gal}$ & PA & $\epsilon$ & \\ [5pt]
% \multicolumn{1}{c}{(1)} & (2) & (3) & (4) & (5) & (6) & (7) & (8) & (9) & (10) & (11) \\
\hline	
FGS02 	& 404 & $178^{+1}_{-2}$ & $0.13^{+0.05}_{-0.03}$ \\ [5pt]
FGS03 	& 37 & $108^{+10}_{-21}$ & $0.31^{+0.11}_{-0.07}$ \\ [5pt]
FGS14 	& 178 & $52^{+8}_{-10}$ & $0.18^{+0.05}_{-0.04}$ \\ [5pt]
FGS17 	& 70 & $98^{+27}_{-42}$ & $0.09^{+0.07}_{-0.02}$ \\ [5pt]
FGS20 	& 57 & $97^{+31}_{-32}$ & $0.09^{+0.08}_{-0.00}$ \\ [5pt]
FGS23 	& 114 & $26^{+16}_{-11}$ & $0.15^{+0.09}_{-0.04}$ \\ [5pt]
FGS26 	& 86 & $19^{+14}_{-15}$ & $0.16^{+0.09}_{-0.06}$ \\ [5pt]
FGS27 	& 191 & $70^{+6}_{-7}$ & $0.24^{+0.06}_{-0.04}$ \\ [5pt]
FGS29 	& 60 & $122^{+8}_{-12}$ & $0.24^{+0.10}_{-0.06}$\\ [5pt]
FGS30 	& 195 & $84^{+9}_{-7}$ & $0.21^{+0.06}_{-0.04}$\\ [5pt]
FGS32 	& 31 & $64^{+9}_{-17}$ & $0.32^{+0.13}_{-0.05}$\\ [5pt]
FGS34 	& 116 & $142^{+26}_{-30}$ & $0.05^{+0.04}_{-0.00}$\\ [5pt]
\hline

\end{tabularx}
\end{table}

For the subsample of 5 FGs we defined in Sect. \ref{1D} we were also able to apply the DS test. We found a significative substructure at c.l. $> 95$\% in FGS02 and FGS27. This means that 40\% of FGs (2 out of 5) have substructures according to DS test. If we considered the analysis of FGS10 carried out in \citet{Aguerri2011}, the fraction decreases to 33\%. This fraction can be compared to other studies in the literature, since the adopted methodology and c.l. are widely used and the number of member galaxies per cluster is roughly the same in all the publications we use for comparison. For example, \citet{Dressler1988} found that 8 out of 14 (corresponding to 57\%) galaxy clusters in their sample show significant subclustering. \citet{Biviano2002} found that 9 out of 23 clusters (corresponding to 39\%) taken from the ESO Nearby Abell Cluster Survey (ENACS) show substructures. Also \citet{Aguerri2010} found a fraction of clusters with substructure of 33\%. Thus, we can conclude that no statistical difference is found between the fraction of FGs and regular clusters that have substructures. It is worth noticing again that our sample is small and that larger ones are needed to prove this result.

With respect to the peculiar velocity of the BGG, only that of FGS14 presents a detectable peculiar velocity in our sample. Also, the BGG of FGS10 shows this peculiar velocity \citep{Aguerri2011}, thus we can estimate a fraction of 33\% for FGs (2 out of 6). These results can be directly compared with that of \citet{Bird1994}, who found that 32\% of the BGGs of their sample clusters has a peculiar velocity. However, the observed discrepancy for FGS10 and FGS14 (about 0.3 and 0.5 times the system $\sigma _v$, respectively) is not particularly high compared to non-FGs \citep{Coziol2009}.

\citet{Burgett2004} applied several tests to a sample of 25 low-richness clusters, including the DS and other velocity tests. They found that 21 clusters show a substructure for at least one of the performed tests. They concluded that substructures are ubiquitous in low-richness clusters. A similar conclusion holds for our FGs, as can be seen from Table \ref{summary}.

Finally, \citet{Miller2012} studied qualitatively the relaxation status in a sample of 12 fossil systems using X-ray observations. They concluded that a fraction between 10\% to 40\% of their fossil systems show substructures. Thus, our results seem to be in agreement with their analysis.

\subsection{Implications for the formation scenario}
The results summarized in Table \ref{summary} show that a significant fraction of FGs have substructures. This result is unexpected, since simulations show that FGs formed at high redshift \citep[half of their mass assembled at $z > 1$,][]{DOnghia2004} and they have enough time to merge all the $M^\ast$ galaxies to form the BGG \citep{DOnghia2005}. Moreover, \citet{vonBenda-Beckmann2008} claimed that FGs accreted less galaxies from the large-scale structure than regular groups/clusters in recent times. Thus, according to these simulations, these old systems had a fast evolution and they should be relaxed at present time. These results were confirmed by \citet{Miraghaei2014} using radio observations of a sample of FGs. In fact, they conclude that the large-scale behavior of FGs is consistent with FGs' relaxed and virialized nature. However, the sample of FGs analyzed in this work shows a large amount of substructures, thus challenging the idea that FGs are old and dynamically relaxed. A mass bias could be the responsible of the observed differences in the fraction of FGs with substructures, since more massive systems should be dynamically younger and have a larger relaxation time \citep{Raouf2014}. In this sense, these authors suggest that only group-mass fossil systems are expected to be dynamically old systems. Using their dating method, the majority of our spectroscopically-confirmed FGs are located in a region of the magnitude of the BGG versus $\Delta m_{12}$ space in which the fraction of young systems vary between 20\% and 50\%. Thus, it is not surprising to find a significant amount of substructures within this sample, according to \citet{Raouf2014}. \citet{Harrison2012} suggested that the evolution of massive FGs is different with respect to group-mass FGs. In particular, they claimed that massive FGs could be the result of the merging between a group-mass FG with another group of galaxies. In this hypothesis, the magnitude gap remains unaffected, but the merger lengthens the relaxation time of the system. It is worth noticing that the subsample of 5 FGs for which we were able to use velocities as a probe for substructures is dominated by massive FGs (M $>10^{14}$ M$_\odot$), so this could be an explanation for the observed large amount of substructures detected using velocities. For group-mass FGs, the dominant mechanism of the evolution is expected to be galaxy-galaxy mergers, with a shorter relaxation time. Unfortunately, our spectroscopy is not enough extended so that we can obtained more than 30 member galaxies in group-mass FGs. For this reason, we can not study the substructures in these systems using velocity information. Moreover, the comparison between the fraction of substructures in fossil and non-fossil groups is not easy due to the lack of systematic studies of substructures in galaxy groups found in the literature.

%%%%%%%%%%%%%%%%%%%%%%%%%%%%%%%%%%
%%%%%%%%%%%%% SECTION 5%%% %%%%%%%%%%%%
%%%%%%%%%%%%%%%%%%%%%%%%%%%%%%%%%%
\section{Conclusions}
\label{conclusions}

We analyzed a sample of 12 spectroscopically-confirmed FGs using a battery of tests that work in the coordinate and/or velocity spaces. The results can be summarized as follows:

\begin{itemize}
\item We applied 1D tests to a subsample of 5 FGs with at least 30 confirmed members. We found that 4 out 5 FGs show hints of substructures in at least one of these tests, the only exception being FGS27.
\item We applied to the same subsample the DS test, which combines velocity and spatial information. Two systems (FGS02 and FGS27) show substructures at a c.l. $\ge 95$\%.
\item The 2D analysis performed using DEDICA and VTP was applied to the whole sample of 12 FGs out to $R_{200}$. Both methods found substructures in 4 systems. The 2D-DEDICA found substructures in FGS02, FGS14, FGS27, and FGS34, whereas the VTP in FGS02, FGS26, FGS27, and FGS30.
\item The ellipticity test gives positive results for the presence of substructures in 2 out of 12 systems (FGS27 and FGS30).
\item We also analyzed the BGGs of all the 12 FGs by looking for signatures of ongoing interactions and mergers. We did not find any signs of recent accretions in these galaxies. However, we were able to apply this test also to FGS28, the smallest and poorest amongst \citet{Santos2007} systems. This system is not part of our sample (see Sect. \ref{sample}) but, interestingly, is the only system that shows two clear tails, probably due to the disruption of two satellite galaxies.
\item None of the FGs in our sample gave positive results in all the applied tests. This confirms that the application of several tests is necessary when looking for substructures in clusters.
\end{itemize}

Although the number of FGs we analyzed is small, we can conclude that FGs show clear signs of substructures. This result seems to discard the hypotheses that all FGs are old and dynamically relaxed. However, a possible mass bias could be an explanation to our result. In particular, the subsample of 5 FGs for which we were able to use velocities as a probe for substructure is composed by massive systems alone (M $> 10^{14}$ M$_\odot$). These massive FGs could have a different evolution with respect of group-mass FGs. In particular, they are suggested to form via the merging of a fossil group with another group of galaxies (whereas the dominant mechanism for the evolution of group-mass FGs is expected to be galaxy-galaxy mergers). This interaction extends the duration of the relaxation process, and for this reason cluster-mass FGs present a similar amount of substructures as non-fossil clusters. In the low-mass regime, we were not able to use velocities to detect substructure. If we consider only the two-dimensional results for the sample of 12 FGs of all masses, we are also unable to detect statistically-robust differences between cluster- and group-mass FGs.  For these reasons, a larger sample of FGs and a larger number of velocities are needed in order to compute statistically-robust fractions and to compare them with that of non-fossil systems in both the low- and high-mass regimes.

\begin{acknowledgements}
This work has been partially funded by the MINECO (grant AIA2013-43188-P). MG acknowledges financial support from MIUR PRIN2010-2011 (J91J12000450001).

This article is based on observations made with the Isaac Newton Telescope, Nordic Optical Telescope, and Telescopio Nazionale Galileo operated on the island of La Palma by the Isaac Newton Group, the Nordic Optical Telescope Scientific Association, and the Fundaci\'on Galileo Galilei of the INAF (Istituto Nazionale di Astrofisica), respectively, in the Spanish Observatorio del Roque de los Muchachos of the Instituto de Astrof\'isica de Canarias.

Funding for the Sloan Digital Sky Survey (SDSS) and SDSS-II has been provided by the Alfred P. Sloan Foundation, the Participating Institutions, the National Science Foundation, the U.S. Department of Energy, the National Aeronautics and Space Administration, the Japanese Monbukagakusho, and the Max Planck Society, and the Higher Education Funding Council for England. The SDSS Web site is http://www.sdss.org/.
The SDSS is managed by the Astrophysical Research Consortium (ARC) for the Participating Institutions. The Participating Institutions are the American Museum of Natural History, Astrophysical Institute Potsdam, University of Basel, University of Cambridge, Case Western Reserve University, The University of Chicago, Drexel University, Fermilab, the Institute for Advanced Study, the Japan Participation Group, The Johns Hopkins University, the Joint Institute for Nuclear Astrophysics, the Kavli Institute for Particle Astrophysics and Cosmology, the Korean Scientist Group, the Chinese Academy of Sciences (LAMOST), Los Alamos National Laboratory, the Max-Planck-Institute for Astronomy (MPIA), the Max-Planck-Institute for Astrophysics (MPA), New Mexico State University, Ohio State University, University of Pittsburgh, University of Portsmouth, Princeton University, the United States Naval Observatory, and the University of Washington.
\end{acknowledgements}

%\bibliographystyle{aa}
%\bibliography{bibliografia}

\end{document}